\documentclass[a4paper,fleqn,usenatbib]{mnras}

\usepackage{newtxtext,newtxmath}
\usepackage[T1]{fontenc}
\usepackage{ae,aecompl}

\usepackage{graphicx}	% Including figure files
\usepackage{amsmath}	% Advanced maths commands
\title[Dust temperature at high redshift]{Analytic models of dust temperature in high-redshift galaxies}

\author[H. Hirashita \& I-D. Chiang]{
Hiroyuki Hirashita\thanks{E-mail: hirashita@asiaa.sinica.edu.tw} and
I-Da Chiang
\\
Institute of Astronomy and Astrophysics, Academia Sinica,
Astronomy--Mathematics Building, No.\ 1, Section 4,
Roosevelt Road, Taipei 10617, Taiwan\\
}

\date{Accepted XXX. Received YYY; in original form ZZZ}

\pubyear{2022}

\begin{document}
\label{firstpage}
\pagerange{\pageref{firstpage}--\pageref{lastpage}}
\maketitle

\begin{abstract}
We investigate physical reasons for high dust temperatures
{($T_\mathrm{dust}\gtrsim 40$ K)}
observed in {some} high-redshift ($z>5$) galaxies
using analytic models. We consider two models that can be treated analytically:
the radiative transfer (RT) model,
{where a broad distribution of values for $T_\mathrm{dust}$ is considered}, and
the one-tempearture (one-$T$) model, which assumes {uniform $T_\mathrm{dust}$}.
These two extremes {serve to bracket the most realistic scenario}.
We adopt the Kennicutt--Schmidt (KS) law to relate
stellar radiation field to gas surface density, and vary the dust-to-gas ratio.
As a consequence, our model is capable of predicting the relation between
the surface density of star formation rate ($\Sigma_\mathrm{SFR}$) or dust mass
($\Sigma_\mathrm{dust}$) and $T_\mathrm{dust}$.
We show that the high $T_\mathrm{dust}$ observed at $z\gtrsim 5$ favour low
dust-to-gas ratios ($\lesssim 10^{-3}$). An enhanced star formation compared with
the KS law gives an alternative explanation for the high $T_\mathrm{dust}$.
The dust temperatures are similar between the two (RT and one-$T$) models as long as we use
ALMA Bands 6--8.
%%Nevertheless, detailed spectral shapes depend on detailed
%%setup of radiative transfer at the dust surface densities observed for high-redshift galaxies.
We also examine the relation among $\Sigma_\mathrm{SFR}$, $\Sigma_\mathrm{dust}$
and $T_\mathrm{dust}$
without assuming the KS law, and confirm the consistency with the actual observational data
at $z>5$.
In the one-$T$ model, we also examine a clumpy dust distribution, which
predicts lower $T_\mathrm{dust}$
because of the leakage of stellar radiation. This enhances the requirement of low dust
abundance or high star formation efficiency to explain the observed high $T_\mathrm{dust}$.
\end{abstract}


%%%%%%%%%%%%%%%%% BODY OF PAPER %%%%%%%%%%%%%%%%%%
\section{Introduction}\label{sec:intro}

The interstellar medium (ISM) of galaxies usually contains dust grains, which play
an important role in various physical processes on galactic or sub-galactic scales.
Dust absorbs and scatters the radiation from stars,
and reradiates it at infrared (IR)--submillimetre (submm) wavelengths
\citep[e.g.][]{Buat:1996aa,Calzetti:2000aa}.
In this way, dust strongly modifies the spectral energy distribution (SED) of interstellar
radiation field and that of galaxy emission
\citep[e.g.][]{Silva:1998aa,Takagi:2003aa,Takeuchi:2005aa}.
This means that a part of star formation activity in a galaxy can only be traced
in the IR--submm \citep[e.g.][]{Kennicutt:1998ab,Inoue:2000aa}, and that
when we extract galaxy properties (stellar mass, age, etc.) from SED fitting,
it is crucial to appropriately consider dust extinction and reemission
\citep[e.g.][]{da-Cunha:2008aa,Boquien:2019aa,Abdurrouf:2021aa,Ferrara:2022aa}.
When dust shields ultraviolet (UV) light, it emits photoelectrons, contributing to the heating of the ISM
\citep[e.g.][]{Tielens:2005aa}.
Dust surfaces are reaction sites for molecular hydrogen formation
\citep[e.g.][]{Gould:1963aa,Cazaux:2004aa}. This makes cold star-forming regions rich in
H$_2$ molecules \citep{Yamasawa:2011aa,Chen:2018aa,Romano:2022aa}.
Dust also induces fragmentation in the star formation process, determining a characteristic mass
of a star. Trough this process, dust also affects the stellar initial mass function (IMF; e.g.\
\citealt{Omukai:2005aa,Schneider:2006aa}).

The above effects of dust may have already been important at high redshift ($z$)
since some galaxies are already dusty at $z>5$
\citep[e.g.][]{Capak:2015aa,Burgarella:2020aa,Fudamoto:2021aa}.
The redshift frontier of dust observation has been expanded to $z>5$
\citep[e.g.][]{Dayal:2018aa} because of the high
capability of the Atacama Large Millimetre/submillimetre Array (ALMA).
For a 'typical' population of high-redshift galaxies, Lyman break galaxies (LBGs), dust emission
has been detected even at $z>7$ \citep[e.g.][]{Watson:2015aa,Laporte:2017aa,Tamura:2019aa,Hashimoto:2019aa,Schouws:2022aa,Inami:2022aa},
although we should also note that most LBGs at such high redshift have too weak
dust emission to be detected by ALMA with a limited time of integration
\citep[e.g.][]{Bouwens:2016aa,Fudamoto:2020aa}.

Observationally, correctly estimating the dust mass is of fundamental importance.
The dust masses derived from ALMA observations for high-redshift galaxies are highly
uncertain because it is difficult to obtain precise dust temperature.
Some studies succeeded in obtaining dust temperatures in LBGs at $z\gtrsim 5$ from
multi-wavelength ALMA data.
A1689-zD, detected with ALMA in Band~6 (1,300~$\micron$;
\citealt{Watson:2015aa}), is later followed up in
Band~7 (870~$\micron$) by \citet{Knudsen:2017aa}, who obtained
a dust temperature of 35--45~K, higher than those in nearby
spiral galaxies ($\sim 20$--25 K; e.g.\ \citealt{Draine:2007aa}).
The high dust temperature of this object
was confirmed by further detections in Band 8 (730~$\micron$; \citealt{Inoue:2020aa}) and Band 9 
(430 $\micron$; \citealt{Bakx:2021aa}).
Including such short wavelengths may be important to trace LBGs with high dust temperatures
\citep{Chen:2022aa}, {as also demonstrated for the above object by \citet{Bakx:2021aa}}.
\citet{Burgarella:2020aa} compiled ALMA detections of LBGs at various $z(>5)$, which enabled
them to statistically trace dust emission SEDs at different restframe wavelengths
\citep[see also][]{Nanni:2020aa,Burgarella:2022aa}, and obtained
dust temperatures of 40--70 K.
\citet{Faisst:2020aa} estimated dust temperatures of four LBGs at $z\sim 5.5$
as 30--43 K \citep[see also][]{Faisst:2017aa}.
\citet{Bakx:2020aa} obtained an even higher dust temperature for a LBG at $z=8.31$ ($>80$ K).
\citet{Sommovigo:2021aa,Sommovigo:2022aa} indirectly derived dust temperatures by utilizing
some empirical relations involving [C \textsc{ii}] 158 $\micron$ emission, obtaining
similar dust temperatures to the above ($\sim 30$--70 K) for a sample of $z>5$ galaxies.
These values imply not only systematically warmer dust than in nearby galaxies but also
a large variety in dust temperature at $z\gtrsim 5$.

The physical reason for high dust temperature is worth clarifying because it may give us a clue to
the evolution of star formation activities and dust properties.
In fact, 
a tendency of increasing dust temperature with redshift is observed at $z\lesssim 4$ 
\citep{Bethermin:2015aa,Schreiber:2018aa,Bethermin:2020aa,Bouwens:2020aa,Faisst:2020aa,Viero:2022aa}, although we need to be careful
about the selection effect \citep{Lim:2020aa}. Cosmological simulations
also predict high dust temperature at high redshift
\citep{Behrens:2018aa,Aoyama:2019aa,Ma:2019aa,Liang:2019aa,Vijayan:2022aa,Pallottini:2022aa}. The tendency of higher dust temperature
at higher redshift could be related to
increasing star formation efficiencies (or equivalently decreasing gas-depletion time-scales;
\citealt{Magnelli:2014aa,Sommovigo:2022aa}).
High dust temperature could also be realized
if star-forming regions have concentrated, compact morphologies 
\citep{Ferrara:2017aa,Behrens:2018aa,Liang:2019aa,Sommovigo:2020aa,Pallottini:2022aa}.
\citet{Sommovigo:2022aa} also considered the effect of dust mass
(as taken into account by other theoretical studies; e.g.\ \citealt{Hirashita:2002aa}) in determining the
dust temperature, which effectively includes {shielding of stellar light}; that is, as the dust mass
increases, the dust shields the stellar radiation and lowers the dust heating per dust mass.
This means that low dust abundance, as well as high stellar radiation intensity, is important for
rising dust temperature towards high redshift.

Since the above conclusions are derived in different contexts,
we here aim at further focusing on the possible essential quantities -- dust abundance
and stellar radiation field -- that affect the dust temperature. This serves to clarify the physical
conditions that could explain the observed high dust temperatures at high redshift.
We formulate the problem by focusing on
physical processes that determine the dust temperature -- heating from stellar radiation and
dust radiative cooling. The balance between these two processes is treated by
an equilibrium condition {as in \citet{Ferrara:2017aa} and \citet{Sommovigo:2020aa}}.
In other words, this paper investigates how the equilibrium condition is affected by the
star formation activities and dust properties.
To make the physical processes transparent, we treat the problem analytically, which
is complementary to some numerical simulations mentioned above.
The transparency of our approach is also useful to
{examine the dust shielding effects with a variety of dust distribution geometries and
dust properties (grain sizes and compositions),}
further serving to examine how robustly dust abundance and stellar radiation field
affect the dust temperature. Utilizing the developed analytic models,
we also address the effects of grain compositions and grain size distribution,
which are suggested to influence the observational properties of galaxies at UV and IR wavelengths
\citep{Yajima:2014aa}.

In this paper, we focus on $z>5$, where the current redshift frontier of dust observation is
located. Nevertheless, we emphasize that the
physical processes treated in this paper are common for any redshift. Thus, the conclusion
drawn this paper is qualitatively applicable to galaxies at $z<5$. In particular, we plan a separate
study for local galaxies by using the framework developed in this paper to further test our
theoretical predictions (Chiang et al., in preparation). Focusing on a certain range of redshift would be
useful to minimize the variation in redshift-dependent
physical processes such as redshift evolution of gas-depletion time \citep{Sommovigo:2022aa},
and systematic difference in stellar populations. 

This paper is organized as follows. We explain the models for dust temperature in Section~\ref{sec:model}.
We show the results in Section~\ref{sec:result}.
We discuss some further issues, especially dependence on various parameters in Section \ref{sec:discussion}. 
Section \ref{sec:conclusion} concludes this paper.
We adopt the following cosmological parameters:
$\Omega_\Lambda=0.7$, $\Omega_\mathrm{M}=0.3$, and
$H_{0}= 70$ km s$^{-1}$ Mpc$^{-1}$.

\section{Model}\label{sec:model}

In this paper, we develop analytic models for dust temperature in a galaxy.
To make an analytic treatment possible, we consider the following two extremes, which
simplify the problem but still catch the essential physical factors affecting the dust temperature:
{(i) one is the case where we consider a distribution of values for the dust temperature,
while (ii) the other assumes a single dust temperature value.}
We refer these models as the (i) \textit{radiative transfer (RT) model} and
(ii) \textit{one-temperature (one-$T$) model},
respectively.
%%Intermediate cases need numerical treatments.
The first model solves radiative transfer {in a simple dust--stars geometry},
while the second could treat another
complexity -- dust distribution geometry. Two models are complementary, and
{catch} different physical aspects that vary the dust temperature under a fixed star
formation activity in a galaxy.

In this paper we do not explicitly consider the effect of the cosmic microwave background
(CMB) on the dust temperature \citep[e.g.][]{da-Cunha:2013aa} since it is redshift-dependent.
Practically, the CMB sets a floor for the dust temperature;
thus, any dust temperature below the CMB temperature, $2.73(1+z)$ K, is not physically
permitted. However, since we are mainly interested in
{galaxies whose observed dust temperatures are}
significantly higher than the CMB temperature, the CMB does not affect our
discussions and conclusions. Thus, we do not apply the redshift-dependent
correction for the CMB temperature so that we do not have to specify the redshift for each result.
Note that the background (including the CMB) is already subtracted from observational data
used for comparison.

\subsection{Basic setup}

\subsubsection{Galaxy properties}\label{subsubsec:galaxy}

We represent the masses of dust, gas, and stars by their surface densities,
denoted as $\Sigma_\mathrm{dust}$, $\Sigma_\mathrm{gas}$ and $\Sigma_\star$,
respectively. A quantity per surface area is convenient since the dust temperature is
determined by the radiation intensity, which has the same dimension as the
surface brightness (luminosity per surface area). For simplicity, we assume that
the dust, gas and stars are distributed in a uniform disc so that the above three quantities
represent the galaxy properties.
Although our formulation implicitly assumes disc geometry, we expect that our results are
not strictly limited to discs because the dust heating radiation in a galaxy has on average
an intensity on the order of $\sim L/(\upi R^2)$, where $L$ and $R$ represent the stellar luminosity
and the optical galaxy size, respectively.
In particular, the geometry factor causes an uncertainty of at most factor 4 ($4\pi$ instead of
$\pi$ in spherical shell geometry; e.g.\ \citealt{Inoue:2020aa}), which affects the
dust temperature only by a factor of $\sim 4^{1/6}\sim 1.26$ at most.
Thus, our results could be applied to any geometry with a 20--30 per cent uncertainty
in the dust temperature.
%%We also assume that the disc is sufficiently thin; that is, the scale height is much smaller than the
%%disc radius.
Since we are interested in normal LBGs, we neglect the contribution
from AGN heating
\citep[see e.g.][for the effect of AGN heating in high-redshift galaxies]{DiMascia:2021aa}.
We also neglect small-scale inhomogeneity that could not be included in our treatment of
smooth surface densities (as commented in Section~\ref{subsec:complex}).
%%and focus rather on global (or spatially averaged) distribution of dust, gas, and stars.

It is empirically established that the SFR is tightly related to the gas mass.
This relation is described by the Kennicutt--Schmidt (KS) law as
\citep{Kennicutt:1998aa}
\begin{align}
\left(\frac{\Sigma_\mathrm{SFR}}{\mathrm{M_{\sun}~yr^{-1}~kpc^{-2}}}\right)=1.0\times 10^{-12}
\kappa_\mathrm{s}\left(\frac{\Sigma_\mathrm{gas}}{\mathrm{M_{\sun}~kpc^{-2}}}\right)^{1.4},
\label{eq:KS}
\end{align}
where $\Sigma_\mathrm{SFR}$ is the surface density of the SFR, and $\kappa_\mathrm{s}$
is the burstiness parameter. Following {\citet{Ferrara:2019aa} and}
\citet{Sommovigo:2021aa}, we include the
correction factor $\kappa_\mathrm{s}$ explicitly, and we adopt $\kappa_\mathrm{s}=1$ for the default
KS law.
We also define the formed stellar mass, $\Sigma_{\star ,0}$, as
\begin{align}
\Sigma_{\star ,0}=\Sigma_\mathrm{SFR}\tau_\star ,\label{eq:Mstar}
\end{align}
where $\tau_\star$ is the age of the star formation activity.
For simplicity, we assume a constant SFR within the duration $\tau_\star$.
The surface luminosity density (stellar luminosity per surface area per frequency $\nu$),
$\mathcal{I}_{\star\nu}$, is calculated as
\begin{align}
\mathcal{I}_{\star\nu}=\Sigma_{\star, 0}\ell_\nu ,
\end{align}
where $\ell_\nu$ is the luminosity density per formed stellar mass, and is calculated
using a spectral synthesis model.
%%With the quantities defined above, we evaluate the dust temperatures based on the
%%two models (i.e.\ RT model and one-$T$ model) formulated below.

To calculate the SED per stellar
mass ($\ell_\nu$), we use
\textsc{starburst99}\footnote{\url{https://www.stsci.edu/science/starburst99/docs/default.htm}}
\citep{Leitherer:1999aa} with a constant SFR and an age of $\tau_\star$.
For simplicity, we fix the stellar SED and assume $\tau_\star\sim 10^8$~yr, which is roughly a typical
stellar age for high-redshift LBGs \citep[][and references therein]{Liu:2019aa}.
Since UV radiation, which saturates in $\sim 10^8$ yr for a constant SFR, is the dominant source of
dust heating \citep{Buat:1996aa},
the resulting dust temperature is not sensitive to the adopted age as long as
$\tau_\star\gtrsim 10^8$ yr.
%%We examine the effect of changing $\tau_\star$ in Section \ref{subsec:age}.
Since we are interested in an early phase of metal enrichment, we set
the stellar metallicity to a sub-solar value, 0.008 ($\sim 1/2$~Z$_\odot$).
{Although we consider a wide range in the dust-to-gas ratio
(note that the gas-phase metallicity is not used in our model),
we fix the stellar metallicity,
since it has much less impact than other parameters (e.g.\ dust-to-gas ratio) on
the dust temperature.}
We adopt the Kroupa initial mass function \citep{Kroupa:2002aa}
with a stellar mass range of 0.1--100 M$_{\sun}$.

\subsubsection{Dust properties}\label{subsubsec:dust_properties}

To make the problem analytically tractable, we neglect scattering, and only consider absorption by dust.
Scattering could raise the chance of absorption because it
effectively increases the path length of the photons. However, the cross-section for scattering
is at most comparable to that of absorption, so that the absorbed energy increases by a factor
of $\sim$2 at most.
%%has a minor influence on the total radiative energy absorbed by dust,
The dust temperature, which depends on the absorbed energy to the power
$\sim 1/6$, does not change significantly. Changing other parameters
such as dust-to-gas ratio, which increases the dust opacity proportionally,
has a larger impact on the dust temperature.
Thus, neglecting scattering does not
influence our conclusions in this paper.

The mass absorption coefficient (absorption cross-section per gas mass),
$\kappa_\mathrm{g,abs}(\nu )$, is evaluated, assuming compact spherical grains, as
\begin{align}
\kappa_\mathrm{g,abs}(\nu )=\mathcal{D}\,
\frac{\int_0^\infty\pi a^2Q_\mathrm{abs}(a,\,\nu )n(a)\,\mathrm{d}a}
{\int_0^\infty\frac{4}{3}\pi a^3sn(a)\,\mathrm{d}a},\label{eq:kappa}
\end{align}
where $\mathcal{D}$ is the dust-to-gas ratio, $a$ is the grain radius,
$Q_\mathrm{abs}(a,\,\nu )$ is
the ratio of absorption to geometrical cross-sections, $s$ is the dust material density,
and $n(a)$ is the grain size distribution, which is defined such that
$n(a)\,\mathrm{d}a$ is the number density
of grains in the radius range from $a$ to $a+\mathrm{d}a$.
The absorption cross-section, specifically $Q_\mathrm{abs}(a,\,\nu )$,
is calculated using the Mie theory \citep{Bohren:1983aa} with silicate or graphite
properties given in \citet{Weingartner:2001aa}.
We adopt $s=3.5$ and 2.24 g cm$^{-3}$ for silicate and graphite, respectively
\citep{Weingartner:2001aa}.

We consider the following power law form for the grain size distribution with index $(-p)$:
\begin{align}
n(a)=
\begin{cases}
        Ca^{-p} & \text{if $a_\mathrm{min}\leq a\leq a_\mathrm{max}$}, \\
        0 & \text{otherwise},
\end{cases}
\end{align}
where $C$ is the normalizing constant.
In this paper, it is not necessary to determine $C$, because it is cancelled out
in $\kappa_\mathrm{g,abs}$ (equation \ref{eq:kappa}).
Since we are not interested in the detailed grain size distribution, we fix
$a_\mathrm{min}=0.001~\micron$ and $a_\mathrm{max}=0.25~\micron$ and only move $p$.
The Milky Way extinction curve can be fitted with $p=3.5$ by mixing silicate and
graphite \citep[][hereafter MRN]{Mathis:1977aa}. Since dust properties at high
redshift is uncertain, we examine silicate and graphite separately. In addition to
$p=3.5$, we also examine $p=2.5$ and 4.5. The shallower (steeper) power $p=2.5$ (4.5)
represents a case where
large (small) grains dominate both dust mass and surface area. Each of these two values
corresponds to
an extreme case where small grains
are efficiently destroyed by sputtering \citep{Hirashita:2015aa}
or produced by shattering \citep{Hirashita:2013ab}.
A larger value of $p$ tends to have a steeper
rise of $\kappa_\mathrm{g,abs}(\nu )$ towards short wavelengths
because of a higher abundance of small grains.
Graphite has a similar mass absorption coefficient to silicate at $\lambda\lesssim 0.15~\micron$
(where $\lambda$ is the rest wavelength),
but has larger values at longer wavelengths.
{The above variations, especially those in $p$, already include
extreme cases for extinction curves, since they produce a larger variety in
the steepness of extinction curves than examined by \citet{Weingartner:2001aa} for
nearby galaxies. Moreover, we will later show that even with those varieties,
variation in grain properties has a minor influence on the dust temperature
(Section \ref{subsec:dust_properties}).}

%%In Fig.\ \ref{fig:kappa}, we show the mass absorption coefficient for the above
%%variation of dust properties. The mass absorption coefficient (per gas mass) is divided
%%by the dust-to-gas ratio to obtain the value per dust mass. We observe that silicate
%%has a rising trend in the UV in general, while
%%graphite has a relatively flat trend with a well known bump at 2175 \AA.

%%\begin{figure}
%%\includegraphics[width=0.48\textwidth]{kappa.eps}
%%\caption{Mass absorption coefficient, $\kappa_\mathrm{g,abs}$, as a function of wavelength
%%$\lambda$.
%%The inverse of the wavelength is shown for the horizontal axis following the convention
%%of extinction curves,
%%and the mass absorption coefficient is divided by the dust-to-gas ratio to present the values
%%per dust mass (independent of the dust-to-gas ratio).
%%The (blue) solid, dashed, and dot--dashed lines show
%%the results for silicate with $p=3.5$ (MRN), 2.5 and 4.5, respectively.
%%The (red) dotted line represents the coefficient for graphite with $p=3.5$.
%%\label{fig:kappa}}
%%\end{figure}

\subsection{Two models}

\subsubsection{RT model}\label{subsubsec:RT}

In this model, we {consider} the multi-temperature effect realized by dust shielding of
stellar light. As mentioned above, we assume homogeneity in the directions parallel to the disc plane
{and that the disc thickness is much smaller than the radial extension of the disc}.
We use coordinate $\zeta$ in the vertical direction with $\zeta =0$
corresponding to the disc mid-plane. {To examine the shielding effect
of dust under the plane-parallel geometry}, we assume that all the
stars are located in the mid-plane (i.e.\ at $\zeta =0$) and that the dust is distributed as `screens'
symmetrically at $\zeta <0$ and $>0$.
{In our treatment, each `layer' of dust has a different dust temperature, so that
multi-dust-temperature structure emerges.}
As mentioned in Section \ref{subsubsec:dust_properties}, we neglect scattering by dust.
With the above setup, we derive the intensity at $\zeta$ on a light path whose direction has an
angle of $\theta$ from the vertical {(positive $\zeta$)} direction.
This intensity (as a function of frequency) is
denoted as
$I_\nu =I_\nu (\zeta ,\,\mu )$, where $\mu\equiv\cos\theta$.

First, we consider UV--optical
wavelengths where
stellar emission is dominant (dust emission is negligible).
The radiative transfer equation including dust absorption and stellar emission is written as
\begin{align}
\mu\frac{\mathrm{d}I_\nu}{\mathrm{d}\zeta}=-\kappa_\mathrm{g,abs}(\nu)\rho (\zeta )
I_\nu+\frac{1}{4\pi}\Sigma_{\star\nu}\delta (\zeta ),\label{eq:radtr}
\end{align}
%%where $\kappa_\mathrm{g,abs}(\nu )$ is the mass absorption coefficient of the gas
%%(dust opacity multiplied by the gas-to-dust mass ratio) at frequency $\nu$, and
where $\rho (\zeta )$ is the gas density at $\zeta$, and $\delta (\zeta )$ is Dirac's
delta function.
%%For simplicity, we neglect scattering, and leave the effect of scattering
%%for future work (Chiang et al., in preparation).
%%This could underestimate the dust optical depth
%%by a factor of $\lesssim 2$; however, dust distribution geometries
%%have a comparable uncertainty, and changing parameters (such as dust-to-gas
%%ratio) have an overwhelming influence on the results.
%%Assuming that the mass absorption coefficient does not depend on $\zeta$,
The above equation can be solved as \citep[see also][]{Hirashita:2019ab}
\begin{align}
I_\nu (\zeta ,\,\mu )=\frac{\mathcal{I}_{\star\nu}}{4\pi\mu}\exp\left(
-\frac{1}{\mu}\kappa_\mathrm{g,abs}(\nu )\,\tilde{\Sigma}_\mathrm{gas} (\zeta )\right) ,
\end{align}
where
\begin{align}
\tilde{\Sigma}_\mathrm{gas}(\zeta )=
\int_0^\zeta\rho (\zeta' )\,\mathrm{d}\zeta' ,\label{eq:Sigma_gas}
\end{align}
is the {surface density measured from the disc mid-plane up to} height $\zeta$.
%%As $\zeta$ approaches the upper edge of the disc, $\tilde{\Sigma}_\mathrm{gas}$
%%approaches $\Sigma_\mathrm{gas}/2$ (the surface density in the upper half of the disc).
In practice,
we use $\tilde{\Sigma}_\mathrm{gas}$ instead of $\zeta$ for
{the integration variable}.
{Since $\mathrm{d}\tilde{\Sigma}_\mathrm{gas}=\rho (\zeta )\,\mathrm{d}\zeta$
(equation \ref{eq:Sigma_gas}),
we do not need to specify the profile of $\rho (\zeta )$ if we use $\tilde{\Sigma}_\mathrm{gas}$
for the integration variable. Thus, we hereafter use $\tilde{\Sigma}_\mathrm{gas}$ not
$\zeta$ to indicate the vertical coordinate.}
We also note that $\tilde{\Sigma}_\mathrm{gas}$ always appears together with
the mass absorption coefficient, so that we could also treat
$\tilde{\Sigma}_\mathrm{dust}\equiv\mathcal{D}\tilde{\Sigma}_\mathrm{gas}$
as {an integration variable} once we give $\mathcal{D}$, which is treated as a constant
parameter in this paper.
{The integration is performed up to a point where
$\tilde{\Sigma}_\mathrm{gas}=\Sigma_\mathrm{gas}/2$ (the total surface density
in the upper plane) is reached.}

The dust temperature at $\tilde{\Sigma}_\mathrm{gas}(\zeta )$,
denoted as $T_\mathrm{dust}(\tilde{\Sigma}_\mathrm{gas})$
is estimated from the radiative equilibrium:
\begin{align}
\int_{912~\text{\AA}}^\infty\kappa_\mathrm{g,abs}(\nu )J_\nu (\tilde{\Sigma}_\mathrm{gas})
\,\mathrm{d}\nu =
\int_0^\infty\kappa_\mathrm{g,abs}(\nu )B_\nu [T_\mathrm{dust}(\tilde{\Sigma}_\mathrm{gas})]\,\mathrm{d}\nu,\label{eq:radeq}
\end{align}
where $J_\nu (\tilde{\Sigma}_\mathrm{gas})$ is the intensity averaged for the solid angle as a function of $\tilde{\Sigma}_\mathrm{gas}$,
and $B_\nu (T_\mathrm{dust})$ is the Planck function at frequency $\nu$ and
dust temperature $T_\mathrm{dust}$.
{The lower limit of the integration range on the left-hand side is set to 912 \AA, since
radiation at shorter wavelengths is mostly absorbed by hydrogen.}
%%As mentioned at the beginning of this section, we do not consider the redshift-dependent CMB effect.
The mean intensity is given by
\begin{align}
J_\nu (\tilde{\Sigma}_\mathrm{gas})=\frac{1}{2}\int_{-1}^1I_\nu (\tilde{\Sigma}_\mathrm{gas},\,\mu )\,\mathrm{d}\mu .
\end{align}
{The main contribution to the integration on the left-hand side of equation (\ref{eq:radeq})
comes from $\lambda\lesssim 4000$ \AA\ \citep[see also][]{Buat:1996aa}, while
that on the right-hand side from IR wavelengths. The numerical integrations are
executed in sufficiently wide wavelength ranges that cover the relevant wavelengths.
We also note that}
$\kappa_\mathrm{g,abs}$ is insensitive to the grain radius
{(or equivalently to the grain size distribution)} at {IR} wavelengths
{since the grain radii are much smaller than the wavelengths}.
Thus, in the real calculation, we use the values and wavelength dependence derived by
\citet{Hirashita:2014aa} when we evaluate the right-hand side of equation (\ref{eq:radeq}) to save
the computational cost; that is,
$\kappa_\mathrm{g,abs}(\nu )=\mathcal{D}\kappa_{158}(\nu /\nu_{158})^\beta$
with $(\kappa_{158},\,\beta )=(13.2~\mathrm{cm^2~g^{-1}},\, 2)$,
$(20.9~\mathrm{cm^2~g^{-1}},\, 2)$ for silicate and graphite, respectively ($\kappa_{158}$
is the dust mass absorption coefficient at $\lambda =158~\micron$, $\nu_{158}$
is the frequency corresponding to $\lambda =158~\micron$, and $\beta$ is the emissivity
index). {This power-law approximation holds for the wavelength range of interest
for dust emission ($\lambda\gtrsim 40~\micron$).} Note that $\mathcal{D}$ is
multiplied to obtain the absorption coefficient per gas mass.
We solve equation (\ref{eq:radeq}) for $T_\mathrm{dust}$ as a function of
$\tilde{\Sigma}_\mathrm{gas}$.

Finally, the dust emission at each layer is superposed to obtain the observed dust SED.
We assume that the dust emission is optically thin, which holds for
the {surface} density range we are interested in (see below).
We calculate the output dust SED per surface area,
$\mathcal{I}_\mathrm{dust}^\mathrm{RT}(\nu )$,
for the RT model as
\begin{align}
\mathcal{I}_\mathrm{dust}^\mathrm{RT}(\nu )=
2\int_0^{\Sigma_\mathrm{gas}/2}\kappa_\mathrm{g,abs}(\nu )(4\upi )B_\nu [T_\mathrm{dust}(\tilde{\Sigma}_\mathrm{gas})]
\,\mathrm{d}\tilde{\Sigma}_\mathrm{gas},
%%\rho (\zeta')\,\mathrm{d}\zeta' ,
\end{align}
%%Don't forget 4pi, since this is the luminosity per physical area (kpc^2), not per solid angle.
where the integration is multiplied by 2 to consider the lower half of the disc.
%%As mentioned above, we practically use $\tilde{\Sigma}_\mathrm{gas}$ for the coordinate
%%(i.e.\ the results do not depend on the density profile in the $\zeta$ direction).
%%We could replace $\rho (\zeta ')\,\mathrm{d}\zeta '$ with $\mathrm{d}\tilde{\Sigma}_\mathrm{gas}$,
%%and the integration range $[0,\,\infty ]$ for $\zeta '$ with $[0,\,\Sigma_\mathrm{gas}/2]$ for
%%$\tilde{\Sigma}_\mathrm{gas}$.

To confirm the optically thin assumption for dust emission, we estimate the
optical depth in the far-IR (FIR), $\tau_\mathrm{FIR}(\lambda )$, as
$\tau_\mathrm{FIR}(\lambda )=0.069(\kappa_{158}/13.2~\mathrm{cm^2~g^{-1}})
(\lambda /100~\micron )^{-2}
(\Sigma_\mathrm{dust}/10^7~\mathrm{M_{\sun}~kpc^{-2}})$. Since we are interested in
the wavelength range $\lambda\gtrsim 100~\micron$ and dust {surface} density
$\sim 10^7~\mathrm{M_{\sun}~kpc^{-2}}$, the optically thin assumption holds.
We do not discuss {surface} densities
$\Sigma_\mathrm{dust}>10^8~\mathrm{M_{\sun}~kpc^{-2}}$
and leave such a high optical depth regime for future work since we need a fully numerical iterative
framework of
energy balance and radiative transfer.

\subsubsection{One-$T$ model}\label{subsubsec:oneT}

In the one-$T$ model, we assume that the radiation field is uniform. This is
an opposite extreme to the RT model in which non-uniformity of the dust temperature naturally emerges.
We basically follow the treatment described by \citet{Inoue:2020aa}
{\citep[see also][for a recent application to high-$z$ galaxies]{Fudamoto:2022aa}}.
%%For comparison with the RT model, we adopt the same
%%$\Sigma_\mathrm{SFR}=0.16$ M$_{\sun}$ yr$^{-1}$ kpc $^{--2}$
%%(or $\Sigma_\mathrm{gas}=10^9$ M$_{\sun}$ kpc$^{-2}$ through the KS law).
Because of the uniformity, we assume that the stars and dust are mixed homogeneously
on a galactic scale. We consider two cases: one is the \textit{homogeneous geometry}, in which
the distribution of dust is smooth and homogeneous, and the other is the \textit{clumpy geometry}, which
allows for clumpy distribution of dust (but the spherical clumps, which have the same radius and density,
are distributed homogeneously). The stars are assumed to be distributed uniformly in both geometries.

To evaluate the dust temperature, we use equation (\ref{eq:radeq}), but
we adopt the following estimate for the radiation field,
$J_\nu =J_\nu^\mathrm{one}$ (note that $J_\nu$ does not depend on the position in the
galaxy by assumption). We denote the escape fraction of the stellar radiation at frequency $\nu$ as
$P_\mathrm{esc}(\tau_\nu )$, where $\tau_\nu$ is the effective optical depth of the dust
at $\nu$. Since the stellar radiation that does not escape from the galaxy is absorbed by dust,
we can relate the escape fraction and $\Sigma_\mathrm{gas}\kappa_\mathrm{g,abs}(\nu )J_\nu^\mathrm{one}$
(absorbed stellar radiation luminosity per surface area) as
\begin{align}
4\upi\Sigma_\mathrm{gas}\kappa_\mathrm{g,abs}(\nu )J_\nu^\mathrm{one}=
\left[ 1-P_\mathrm{esc}(\tau_\nu )\right]\mathcal{I}_{\star\nu} ,\label{eq:radeq_one}
\end{align}
We evaluate $P_\mathrm{esc}(\tau_\nu )$ using the averaged escape fraction in
a plane-parallel disc, where we use the optical depth in the vertical direction
for $\tau_\nu$ (evaluated below).
The averaged escape fraction of a plane-parallel disc in the direction
which has an angle $\theta$ ($0\leq\theta <\upi /2$) from the vertical direction is
$(1-\mathrm{e}^{-\tau_\nu /\mu})/(\tau_\nu /\mu )$ . Thus, averaging it over all the solid angle,
we obtain the escape fraction as
\begin{align}
P_\mathrm{esc}(\tau_\nu )=\int_0^1\frac{1-\mathrm{e}^{-\tau_\nu /\mu}}{\tau_\nu /\mu}\,
\mathrm{d}\mu .\label{eq:Pesc}
\end{align}

{The optical depth $\tau_\nu$ is estimated in different ways for
the homogeneous and clumpy geometries as explained below.}

\paragraph*{Homogeneous geometry}

For the homogeneous geometry, the effective optical depth
$\tau_\nu =\tau_\nu^\mathrm{hom}$ is simply evaluated as
\begin{align}
\tau_\nu^\mathrm{hom}=\kappa_\mathrm{g,abs}(\nu )\Sigma_\mathrm{gas}.
\end{align}
Note that we evaluate the optical depth for plane-parallel geometry
(while \citealt{Inoue:2020aa} adopted
spherical geometry).

\paragraph*{Clumpy geometry}

For the clumpy geometry, $\tau_\nu =\tau_\nu^\mathrm{cl}$, which is given below.
We adopt the high-contrast limit, in which the clumps are much denser than the
interclump medium, since the case with low contrast is similar to the homogeneous case.
As shown by \citet{Inoue:2020aa},
\begin{align}
\tau_\nu^\mathrm{cl}=\tau_\nu^\mathrm{hom}P_\mathrm{esc}^\mathrm{sp}
(\tau_\mathrm{c,\nu}),
\end{align}
where $\tau_\mathrm{c,\nu }$ is the radial optical depth of a single clump at $\nu$,
and $P_\mathrm{esc}^\mathrm{sp}(\tau_\mathrm{c,\nu})$
is the escape fraction of a homogeneous sphere
of optical
depth $\tau_\nu^\mathrm{cl0}$ given by \citep{Varosi:1999aa,Osterbrock:2006aa}
\begin{align}
P_\mathrm{esc}^\mathrm{sp}(\tau_\mathrm{c,\nu})=\frac{3}{4\tau_\mathrm{c,\nu}}
\left[ 1-\frac{1}{2{\tau_\mathrm{c,\nu}}^2}+
\left(\frac{1}{\tau_\mathrm{c,\nu}}+\frac{1}{2{\tau_\mathrm{c,\nu}}^2}\right)
\mathrm{e}^{-2\tau_\mathrm{c,\nu}}\right] .
\end{align}
In the high-contrast approximation,
$\tau_\nu^\mathrm{c,\nu}\simeq\tau_\nu^\mathrm{hom}\xi_\mathrm{cl}$,
where $\xi_\mathrm{cl}$ is the clumpiness parameter:
The limit $\xi_\mathrm{cl}\to 0$ corresponds to an infinite number
of infinitely compact clumps (reduced to the homogeneous case),
while the opposite $\xi_\mathrm{cl}\to\infty$ means a small number of infinitely
compact clumps (reduced to no absorption; \citealt{Inoue:2020aa}).
An intermediate value of $\xi_\mathrm{cl}$ is,
thus, interesting for the clumpy geometry.
%%This parameter roughly gives the inverse of the projected
%%filling factor of the clumps if $\xi_\mathrm{cl}>1$.
%%\citet{Inoue:2020aa} found that $\xi_\mathrm{cl}\sim 0.1$ fits the dust emission
%%SED of a high-redshift galaxy at $z=7.5$ \citep[A1689zD1;][]{Watson:2015aa}.
%%We use this value as a fiducial value, and examine a range of $\xi_\mathrm{cl}=0.01$--1.

\medskip

The overall procedures are summarized here.
We adopt $\tau_\nu =\tau_\nu^\mathrm{hom}$ for the homogeneous geometry
or $\tau_\nu =\tau_\nu^\mathrm{cl}$ for the clumpy geometry to evaluate the escape fraction,
$P_\mathrm{esc}(\tau_\nu )$ in
equation~(\ref{eq:Pesc}).
Using $P_\mathrm{esc}$, we obtain $J_\nu^\mathrm{one}$ in equation (\ref{eq:radeq_one}).
This $J_\nu^\mathrm{one}$ is then used in equation~(\ref{eq:radeq}) by replacing
$J_\nu (\tilde{\Sigma}_\mathrm{gas} )$ with
$J_\nu^\mathrm{one}$. Solving this equation for
%%$T_\mathrm{dust}=T_\mathrm{dust}^\text{one-$T$}$,
$T_\mathrm{dust}$, we obtain the dust temperature
in the one-$T$ model. We simply denote the dust temperature in the one-$T$ model as $T_\mathrm{dust}$.
Assuming optically thin emission, the dust emission SED per surface area,
$\mathcal{I}_\mathrm{dust}^\text{one-$T$}(\nu )$,
in this model is estimated as
\begin{align}
\mathcal{I}_\mathrm{dust}^\text{one-$T$}(\nu )=4\upi\kappa_\mathrm{g,abs}(\nu )\Sigma_\mathrm{gas}
B_\nu (T_\mathrm{dust}) .
\end{align}

\subsection{Colour temperature}\label{subsec:clr}

Observationally, dust temperature is derived from multi-wavelength observations at rest-frame
FIR wavelengths. Thus, observationally estimated dust temperatures are basically colour
temperatures.
The colour temperature is defined using the intensities at two wavelengths,
$\lambda_1$ and $\lambda_2$
(corresponding frequencies $\nu_1$ and $\nu_2$, respectively),
and is denoted as $T_\mathrm{clr}(\lambda_1,\,\lambda_2)$.

For the RT model, the colour temperature is obtained by solving the following
equation for $T_\mathrm{clr}$ \citep[e.g.][]{Krugel:2003aa}:
\begin{align}
\frac{\kappa_\mathrm{g,abs}(\nu_2)B_{\nu_2} [T_\mathrm{clr}(\lambda_1,\,\lambda_2)]}
{\kappa_\mathrm{g,abs}(\nu_1)B_{\nu_1} [T_\mathrm{clr}(\lambda_1,\,\lambda_2)]}=
\frac{\mathcal{I}_\mathrm{dust}^\mathrm{RT}(\nu_2)}{\mathcal{I}_\mathrm{dust}^\mathrm{RT}(\nu_1)}.
\end{align}
The colour temperature depends on the choice of $\lambda_1$ and
$\lambda_2$.
Strictly speaking,
this colour temperature is not really the one derived from the observation, since
we do not know the real frequency dependence of $\kappa_\mathrm{g,abs}$ in the observed
galaxy.
We need to keep in mind a larger uncertainty caused by the unknown frequency dependence of
the mass absorption coefficient in actual observations, but we neglect it in this paper.
We basically take $\lambda_1=100~\micron$ and $\lambda_2=200~\micron$, so that the
two wavelengths are near to ALMA bands (Bands 6, 7, and 8) for galaxies at $z>5$.
{A frequently adopted choice $\lambda_1 =88~\micron$ and
$\lambda_2=158~\micron$ tuned to the [O \textsc{iii}] and [C \textsc{ii}] emissions,
respectively, have similar results so that the following results holds for this choice.
We further} examine the
effects of wavelength choice in Section \ref{subsec:radtr_vs_oneT}.
We also comment on a caution in using the 450 $\micron$ band (Band 9) for high-redshift galaxies there.

For the one-$T$ model, the dust temperature has only a single value. Thus,
$T_\mathrm{clr}(\lambda_1,\,\lambda_2)=T_\mathrm{dust}$ always holds in the
one-$T$ model.

\subsection{Observational data for comparison}\label{subsec:sample}

\begin{table*}
\caption{High-redshift galaxies used for comparison.}
\label{tab:sample}
\begin{tabular}{lccccccccc}
\hline
Name & $z$ & $T_\mathrm{dust}$ & $L_\mathrm{UV}$ &
$L_\mathrm{IR}$ & $M_\mathrm{dust}$ & $\sqrt{ab}/2$ &
$\Sigma_\mathrm{SFR}$ &
$\Sigma_\mathrm{dust}$ & ref.$^a$\\
 & & (K) & ($10^{11}$\,L$_{\sun}$)& ($10^{11}$\,L$_{\sun}$) & ($10^{7}$\,M$_{\sun}$) & (kpc) &
 (M$_{\sun}$\,yr$^{-1}$\,kpc$^{-2}$) & ($10^7$\,M$_{\sun}$\,kpc$^{-2}$) & \\
\hline
{A2744\_YD4} & 8.38 & $>55$ & 0.25 & $>1.8$ & $<0.18$ & 0.50$^b$ & $>36$ & $<0.23$
& 1, 2\\
MACS0416\_Y1 & 8.31 & $>80^b$ & 0.45 & $>11.1^b$ & $<0.035^c$ & 0.45 & $>250$ & $<0.056$
& 3, 4\\
B14-65666 & 7.15 & 30--80$^d$ & 2.0 & 3--30$^c$ & 0.5--30$^d$ & $0.87\pm 0.30$  &
$65^{+109}_{-38}$ & $1.7^{+11.6}_{-1.5}$ & 5, 6\rule[0mm]{0mm}{3mm}\\
A1689-zD1 & 7.13 & $43^{+13}_{-7}$ & 0.18 & $1.9^{+0.5}_{-0.4}$ & $1.7^{+1.3}_{-0.7}$
& $0.77\pm 0.10$ & $14^{+3}_{-3}$ & $0.91^{+0.70}_{-0.38}$ & 7, 8\rule[0mm]{0mm}{3mm}\\
J1211-0118 & 6.03 & $38^{+34}_{-12}$ & 2.7 & $3.2^{+18.7}_{-1.7}$ & $3.0^{+10.5}_{-2.3}$ & 2.0$^e$
& $7.3^{+19.4}_{-1.8}$ & $0.24^{+0.84}_{-0.18}$ & 9\rule[0mm]{0mm}{3mm}\\
J0217-0208 & 6.20 & $25^{+19}_{-5}$ & 4.3 & $1.4^{+2.5}_{-0.3}$ & $19^{+735}_{-16}$ & 2.0$^e$ & $8.5^{+2.7}_{-0.3}$ & $1.57^{+60.8}_{-1.32}$ & 9\rule[0mm]{0mm}{3mm}\\
HZ4 & 5.54 & $57^{+67}_{-17}$ & 1.8 & $8.1^{+10.9}_{-7.1}$ & $1.1^{1.2}_{-0.8}$$^f$ & 0.72$^g$
& $81^{+87}_{-57}$ & $0.65^{+0.76}_{-0.47}$ & 10, 11\rule[0mm]{0mm}{3mm}\\
HZ6 & 5.29 & $41^{+18}_{-7}$ & 2.1 & $5.4^{+3.5}_{-2.9}$ & $4.6^{+3.2}_{-2.5}$$^f$ & 3.36$^g$
& $3.0^{+1.3}_{-1.1}$ & $0.13^{+0.09}_{-0.07}$ & 10, 11\rule[0mm]{0mm}{3mm}\\
HZ9 & 5.54 & $49^{+29}_{-11}$ & 0.85 & $14^{+8.6}_{-8.9}$ & $4.3^{+3.5}_{-2.6}$$^f$ & 0.95$^g$
& $62^{+39}_{-41}$ & $1.5^{+1.2}_{-0.9}$ & 10, 11\rule[0mm]{0mm}{3mm}\\
HZ10 & 5.66 & $46^{+16}_{-9}$ & 2.3 & $31^{+13}_{-14}$ & $14^{+9}_{-7}$ &
$1.50\pm 0.44$ & $57^{+23}_{-25}$ & $2.0^{+1.3}_{-0.9}$ & 10, 11\rule[0mm]{0mm}{3mm}\\
\hline
\end{tabular}
\\
  \begin{minipage}{1.9\columnwidth}
{Note -- corrected for lensing for A2744\_YD4, MACS0416\_Y1, and A1689-zD1
with correction factor $\mu=1.8$, 1.4, and 9.3, respectively.}\\
$^a$References: {1) \citet{Laporte:2017aa}; 2) \citet{Laporte:2019aa};} 3) \citet{Bakx:2020aa}; 
4) \citet{Tamura:2019aa}; 5) \citet{Sugahara:2021aa}; 6) \citet{Hashimoto:2019aa};
7) \citet{Inoue:2020aa}; 8) \citet{Bakx:2021aa}; 9) \citet{Harikane:2020aa}; 10) \citet{Capak:2015aa}:
11) \citet{Faisst:2020aa}.\\
{$^b$We assumed $0\farcs 5\times0\farcs 3$
from fig.\ 1 of \citet{Laporte:2017aa}, and corrected for lensing.}\\
$^c$The results for $\beta =2$ are adopted.\\ %%with a correction factor of $\mu=1.43$ for lensing.\\
$^d$The results for modified blackbody fitting with $\beta =2$ are adopted.\\
$^e$We adopt the diameter ($0\farcs 7$) used to measure the flux.\\
$^f$Estimated from the ALMA Band 7 flux using the dust temperature given in the literature
(also listed in this table) and the silicate
mass absorption coefficient used in this paper. We estimate the error based on the uncertainty
in the dust temperature, which is dominant in the error budget.\\
$^g$Radius in the rest-frame UV (not spatially resolved by ALMA).
  \end{minipage}
\end{table*}

We compare the results with observational data at $z>5$, which are listed
in Table \ref{tab:sample}.
We selected galaxies with dust temperature measurements
from multi-wavelength ALMA {observations}, as compiled by
\citet[][see their fig.~4]{Bakx:2021aa}.
We do not include indirect measurements through [C \textsc{ii}] 158 $\micron$
emission \citep{Sommovigo:2022aa} or with the help of UV optical depth \citep{Ferrara:2022aa};
these methods show a consistent range of
dust temperature ($\sim 40$--60~K).
The SFR is evaluated based on the UV and IR luminosities, which trace unobscured and
obscured star formation activity, respectively. The IR luminosity, $L_\mathrm{IR}$, is the
integrated luminosity in wavelength range 3--1000 $\micron$,
and the UV luminosity, $L_\mathrm{UV}$ is estimated by $\nu L_\nu$ (luminosity density multiplied by
the frequency)
at a rest-frame wavelength in the range of 0.15--0.2 $\micron$ (in this range the exact choice of
$\lambda$ does not affect the results significantly).
We obtain the SFR as
$\mathrm{SFR}=C_\mathrm{UV}L_\mathrm{UV}+C_\mathrm{IR}L_\mathrm{IR}$,
where we adopt conversion coefficients
{$C_\mathrm{UV}=2.0\times 10^{-10}$ M$_{\sun}$ yr$^{-1}$ L$_{\sun}^{-1}$  %3.9
and $C_\mathrm{IR}=1.3\times 10^{-10}$ M$_{\sun}$  yr$^{-1}$ L$_{\sun}^{-1}$ %1.8
derived from the stellar SED at $t=10^8$ yr adopted in this paper (Section \ref{subsubsec:galaxy})
based on the method described by \citep{Hirashita:2003aa}.
These coefficients may change at most by a factor of 2 if a different stellar age is adopted
(30--300 Myr); however, the change is smaller than the errors in $L_\mathrm{IR}$.}
We evaluate the error in the SFR using the
uncertainty in $L_\mathrm{IR}$, which is dominant over that in $L_\mathrm{UV}$
mainly because of the uncertainty in the dust temperature.
For the dust mass ($M_\mathrm{dust}$), we confirmed
that the adopted mass absorption coefficient in the literature is consistent with our silicate value
within the uncertainty caused by the dust temperature.
To derive surface densities, we also need the surface area, which is evaluated by
$\upi ab$, where $a$ and $b$ are the semi-major and semi-minor axis of the physical size, respectively.
We list $\sqrt{ab}/2$ for the mean radius in the table. Unless otherwise stated in the note,
we adopt the size measurements of $\sqrt{ab}$ from ALMA.
SFR and $M_\mathrm{dust}$ are divided by $\upi ab$ to obtain $\Sigma_\mathrm{SFR}$ and
$\Sigma_\mathrm{dust}$.

We note that the surface densities in the models
($\Sigma_\mathrm{dust}$ and $\Sigma_\mathrm{SFR}$)
are measured in the vertical direction of the disc, whereas the observational data are not corrected for the
inclination. However, the correction factor is $\sim 2$ on average
(based on the average of $\cos\theta$ in
all the solid angle), while the error bars are even larger. Moreover, a factor 2 shift of the observational
data does not change the discussions and conclusions in this paper. Therefore, we neglect the inclination
effects in comparison with observational data.

\section{Results}\label{sec:result}

We show how the dust temperatures in the two (RT and one-$T$) models
are affected by various galaxy parameters.
We display the dust (colour) temperature as a function of surface densities, for which we
adopt $\Sigma_\mathrm{SFR}$ and $\Sigma_\mathrm{dust}$.
The first quantity regulates
the radiation field intensity, while the second directly reflects the effect of dust optical depth.
It is useful to remind the reader that these two surface densities are related by the KS law
(equation \ref{eq:KS}) as
\begin{align}
\left(\frac{\Sigma_\mathrm{SFR}}{\mathrm{M_{\sun}~yr^{-1}~kpc^{-2}}}\right)=1.0\times 10^{-12}
\kappa_\mathrm{s}\mathcal{D}^{-1.4}\left(\frac{\Sigma_\mathrm{dust}}{\mathrm{M_{\sun}~kpc^{-2}}}\right)^{1.4}.\label{eq:KS_dust}
\end{align}
Therefore, $\kappa_\mathrm{s}$ and $\mathcal{D}$ are degenerate in such a way that the
same value of $\kappa_\mathrm{s}\mathcal{D}^{-1.4}$ produces the same result.
For example, lowering $\mathcal{D}$ has the same effect as raising $\kappa_\mathrm{s}$.

Since the gas mass is difficult to obtain for high-redshift galaxies, we do not use $\Sigma_\mathrm{gas}$.
However, since $\Sigma_\mathrm{gas}$ has a simple scaling with $\Sigma_\mathrm{SFR}$,
it is easy to convert $\Sigma_\mathrm{SFR}$ to $\Sigma_\mathrm{gas}$ using equation (\ref{eq:KS}).

In this section, we adopt silicate with $p=3.5$ and focus on the variation in the dust abundance
($\mathcal{D}$), and leave the discussion on the variation of dust properties to
Section \ref{subsec:dust_properties}.
In the one-$T$ model, we concentrate on the homogeneous geometry and
we separately discuss the comparison with the clumpy geometry in
Section \ref{subsec:dust_properties}.

\subsection{Relation between SFR and dust temperature}\label{subsec:SFR_Tdust}

We show the dust temperature as a function of SFR surface density.
Note that high $\Sigma_\mathrm{SFR}$ implicitly indicates high
dust {surface} density because of the KS law
(equation \ref{eq:KS_dust}). %%\propto (\Sigma_\mathrm{dust}/\mathcal{D})^{1.4}$.
As mentioned in Section \ref{subsec:clr}, we take the colour temperature at rest-frame
100 and 200 $\micron$ in the RT
model. We also vary
$\mathcal{D}=\Sigma_\mathrm{dust}/\Sigma_\mathrm{gas}=10^{-4}$, $10^{-3}$ and
$10^{-2}$ to examine the effect of dust abundance.
We assume the KS law with $\kappa_\mathrm{s}=1$ by default,
and also examine a bursty star formation with $\kappa_\mathrm{s}=10$.

\begin{figure}
\begin{center}
\includegraphics[width=0.48\textwidth]{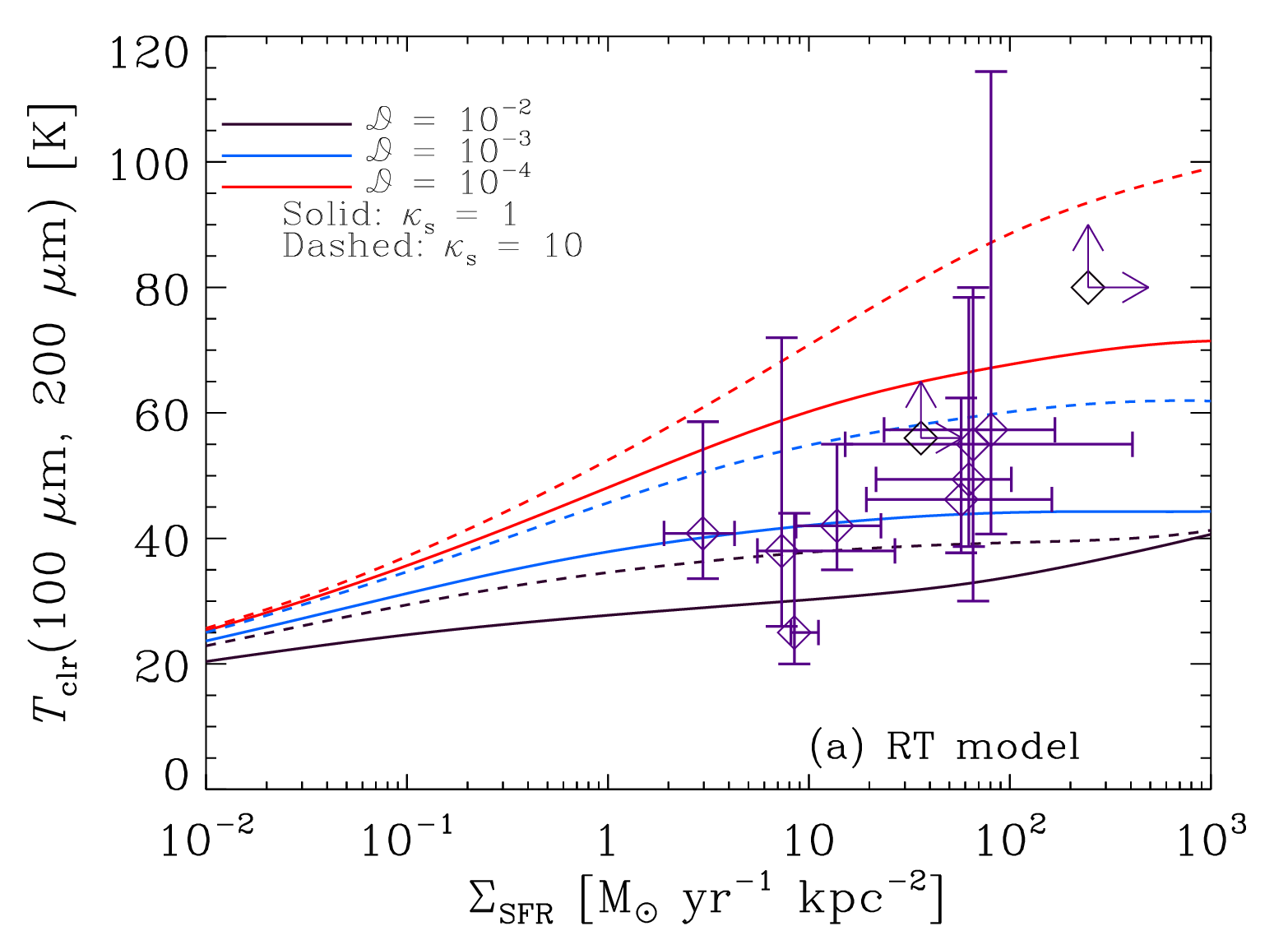}
\includegraphics[width=0.48\textwidth]{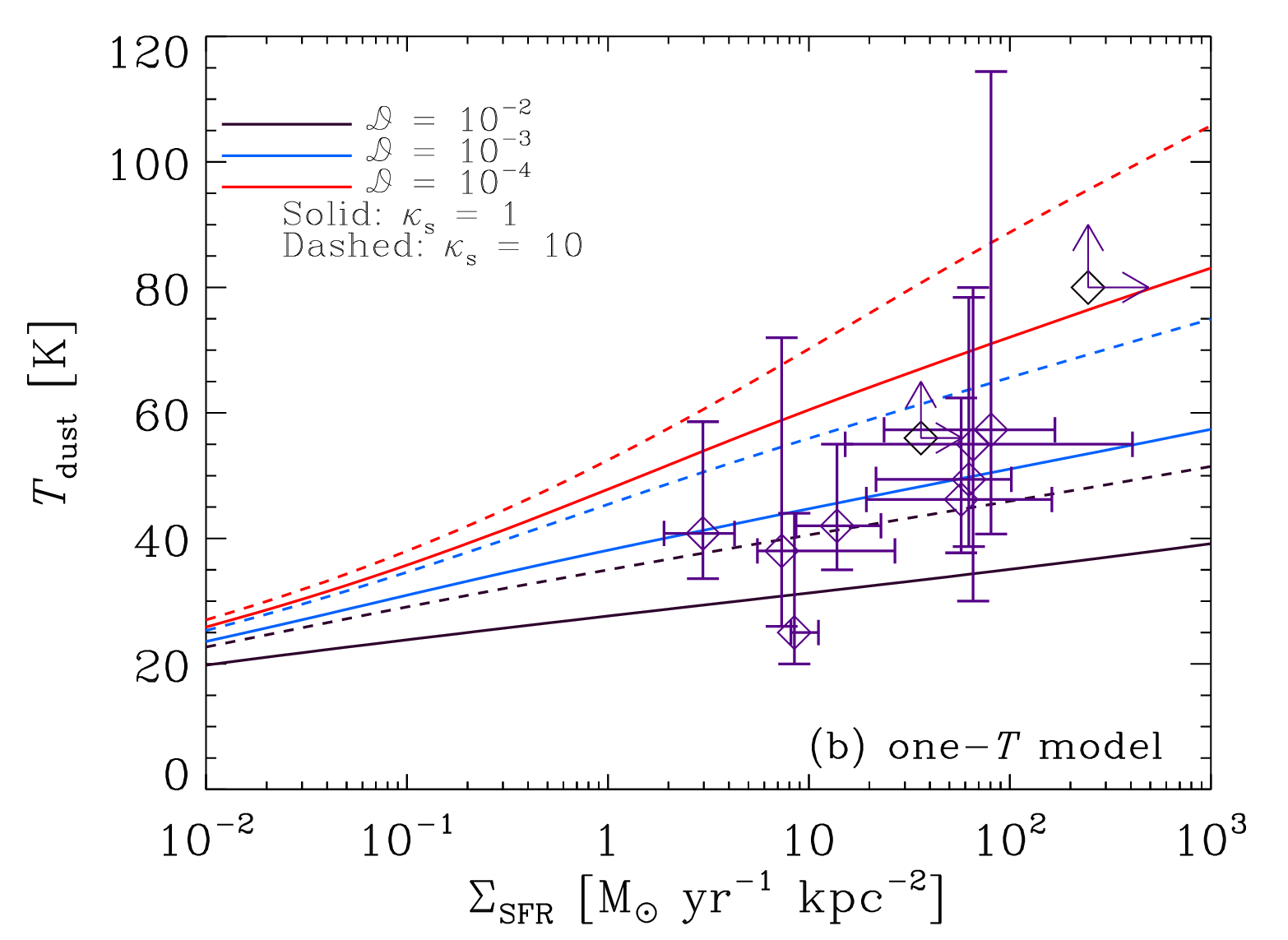}
\end{center}
\caption{(a) Colour temperature as a function of SFR surface density
($\Sigma_\mathrm{SFR}$) in the RT model.
%%Note that $\Sigma_\mathrm{gas}$ shown
%%in Fig.\ \ref{fig:clr_gas} and $\Sigma_\mathrm{SFR}$ presented here are related by the KS law
%%(equation \ref{eq:KS}).
The black, blue, and red lines correspond to $\mathcal{D}=10^{-2}$, $10^{-3}$, and
$10^{-4}$, respectively. The solid and dashed lines show the results for $\kappa_\mathrm{s}=1$
(KS law) and for $\kappa_\mathrm{s}=10$ (a burst of star formation), respectively.
(b) Dust temperature as a function of $\Sigma_\mathrm{SFR}$ in the one-$T$ model.
The line species indicate the same meaning as in Panel (a).
In both panels, we also display observational data (points with error bars or with arrows for
lower/upper limits) summarized in Table~\ref{tab:sample}.
\label{fig:clr_SFR}}
\end{figure}

In Fig.\ \ref{fig:clr_SFR}a, we show the color temperature at 100 and 200 $\micron$
as a function of $\Sigma_\mathrm{SFR}$ for the RT model.
%%for various dust-to-gas ratio.
%%We adopt the silicate model with $p=3.5$, and only present the
%%RT model since the one-$T$ model produces similar results as shown above.
The colour temperature rises as the SFR surface density increases, although
high $\Sigma_\mathrm{SFR}$ also indicates high dust {surface} density
(equation \ref{eq:KS_dust}).
This is explained by the following scaling arguments:
The KS law indicates that
$\Sigma_\mathrm{dust}\propto\Sigma_\mathrm{SFR}^{1/1.4}$ while the
increase of radiation field is proportional to $\Sigma_\mathrm{SFR}$. Thus, the increase of
radiation field is more significant than that of dust {surface} density,
which means that the dust temperature rises as $\Sigma_\mathrm{SFR}$ becomes higher.
We also observe that the colour temperature is sensitive to the dust-to-gas ratio, especially
at high $\Sigma_\mathrm{SFR}$. The rise is steeper for smaller $\mathcal{D}$.
The trend of higher dust temperature for lower $\mathcal{D}$ is
{due to the fact
that in a dust poor environment, at fixed $\Sigma_\mathrm{SFR}$, the dust is exposed
to less shielded UV field, thus being more efficiently heated \citep[see also][]{Sommovigo:2022aa}.}

We show the results of the one-$T$ model
%%the dust temperature as a function of $\Sigma_\mathrm{SFR}$
in Fig.~\ref{fig:clr_SFR}b. Recall that
the dust temperature is the same as the colour temperature because of the
single-temperature assumption. Overall, we find similar dust temperatures to those shown for
the RT model. The difference between the two models is much smaller than
the variation caused by different dust-to-gas ratios. 

We also examine a burst mode of star formation, which is realized by raising $\kappa_\mathrm{s}$
(equation \ref{eq:KS}) in our model. For a burst mode, we examine $\kappa_\mathrm{s}=10$,
which is {inferred} for some high-redshift starbursts \citep[e.g.][]{Vallini:2020aa,Vallini:2021aa,Sommovigo:2021aa,Ferrara:2022aa} and is
expected from simulations \citep{Pallottini:2022aa}.
We observe in Fig.\ \ref{fig:clr_SFR} that the dust temperatures are raised by the
burst (i.e.\ higher radiation field). In this sense, raising $\kappa_\mathrm{s}$ has a similar effect to
decreasing $\mathcal{D}$.
This is because of the degeneracy mentioned above (equation \ref{eq:KS_dust}): The same result is obtained
for the same value of $\kappa_\mathrm{s}\mathcal{D}^{-1.4}$.

We also plot the observational data (Section \ref{subsec:sample}; Table \ref{tab:sample})
in Fig.~\ref{fig:clr_SFR}. We find that the data points favour low $\mathcal{D}$ and/or high
$\kappa_\mathrm{s}$, although the large error bars make it difficult to obtain a firm constraint
on these parameters. If the
dust-to-gas ratio is comparable to the value seen in the Milky Way and nearby solar-metallicity galaxies
($\mathcal{D}\sim 0.01$), the dust temperature does not exceed 50 K even with
$\kappa_\mathrm{s}=10$ at as high
$\Sigma_\mathrm{SFR}$. Thus, if the dust temperature
is higher than $\sim$50 K {as observationally indicated for some $z>5$ galaxies}
%%in galaxies at $z>5$ as expected from the extrapolation of the trend
%%at $z\lesssim 4$ \citep{Schreiber:2018aa},
it is highly probable that the dust-to-gas ratio is
significantly lower than the Milky Way value.
It is also interesting to point out that there is a hint of positive correlation between
$\Sigma_\mathrm{SFR}$ and dust temperature in the observational data although the errors are
admittedly large. This positive correlation is consistent with more dust heating radiation in
more actively star-forming galaxies.

\subsection{Relation between dust mass and dust temperature}\label{subsec:Sigma_dust}

\begin{figure}
\begin{center}
\includegraphics[width=0.48\textwidth]{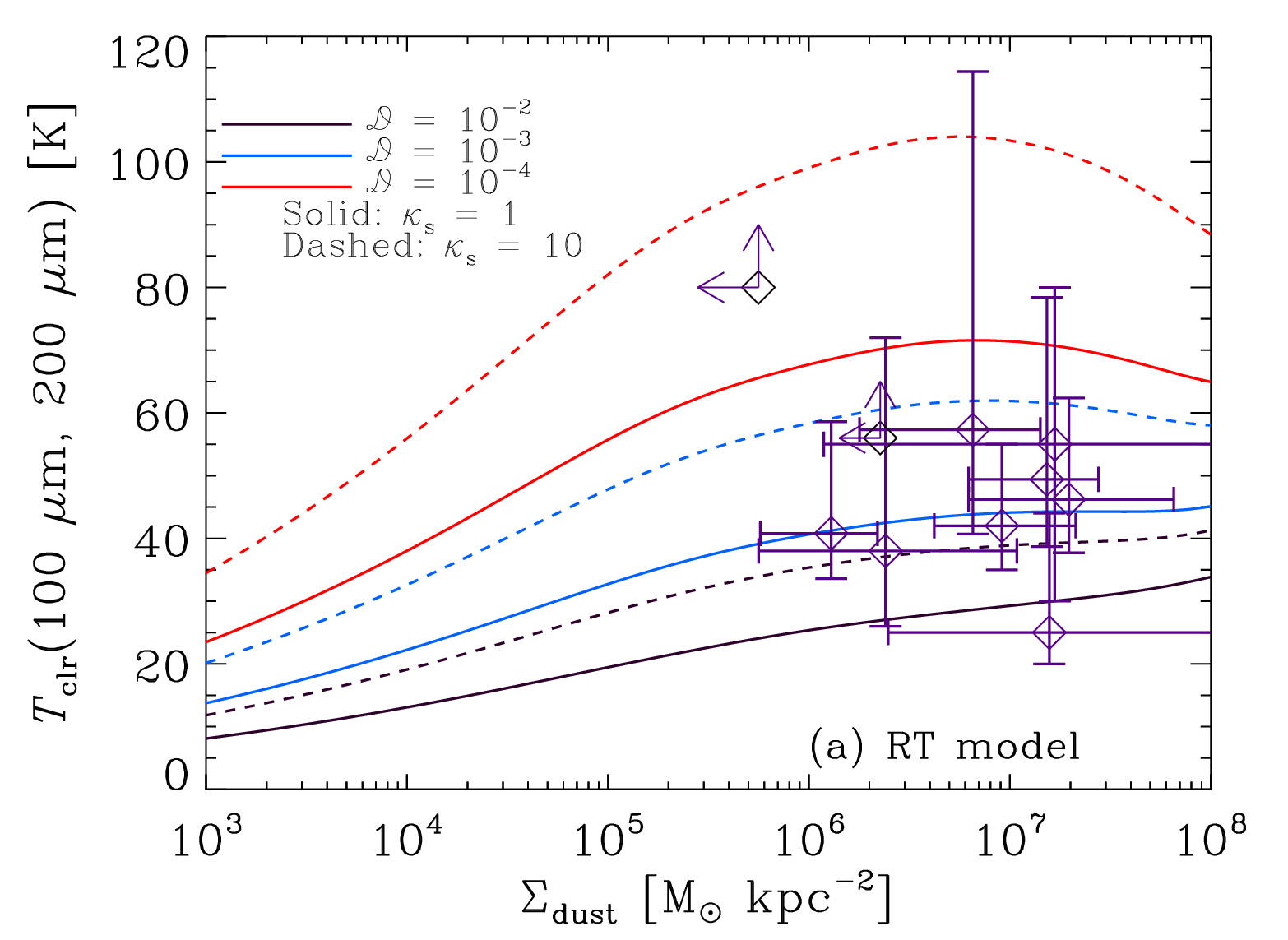}
\includegraphics[width=0.48\textwidth]{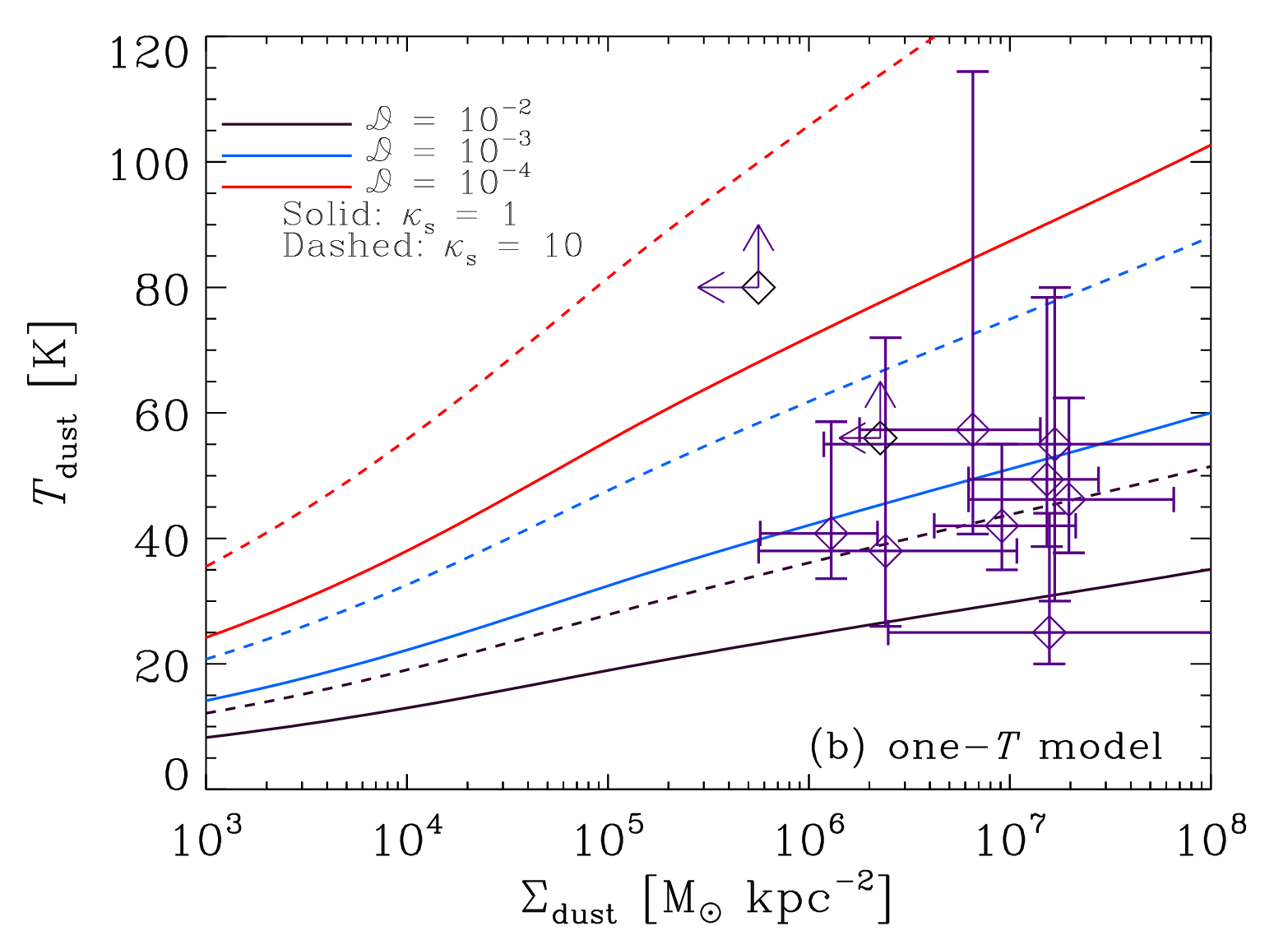}
\end{center}
\caption{(a) Colour temperature as a function of dust surface density
($\Sigma_\mathrm{dust}$) in the RT model.
(b) Dust temperature as a function of $\Sigma_\mathrm{SFR}$ in the RT model.
The line species and data points are the same as in Fig.\ \ref{fig:clr_SFR}.
\label{fig:clr_dust}}
\end{figure}

Dust {surface} density has a direct impact on dust temperature through the shielding
of stellar radiation. Thus, we expect that it is useful to examine the relation between dust
surface density and dust temperature. We show this relation in Fig.\ \ref{fig:clr_dust}.
Note that we only present $\Sigma_\mathrm{dust}$ up to
$10^8$ M$_{\sun}$ kpc$^{-2}$, beyond which the optically thin assumption
for the FIR radiation breaks down (Section \ref{subsubsec:RT}).
We confirm that the observational sample mostly has
$\Sigma_\mathrm{dust}<10^8$ M$_{\sun}$ kpc$^{-2}$.

In Fig.\ \ref{fig:clr_dust}, we observe that the difference in dust temperature among
various dust-to-gas ratios
and burstiness parameters are
clear at all $\Sigma_\mathrm{dust}$. This is because, with a fixed value of $\Sigma_\mathrm{dust}$,
$\Sigma_\mathrm{SFR}$ is higher for lower $\mathcal{D}$ (equation \ref{eq:KS_dust}). A
high value of the burstiness parameter ($\kappa_\mathrm{s}$) also raises the dust temperature;
thus, as noted above, a high burstiness parameter has the same effect as
a low $\mathcal{D}$. Both of the RT and one-$T$ models predict similar dust temperatures at
$\Sigma_\mathrm{dust}\lesssim 10^7$ M$_{\sun}$ kpc$^{-2}$. In the RT model,
the rise of the dust temperature is saturated at
$\Sigma_\mathrm{dust}\sim 10^7$ M$_{\sun}$ kpc$^{-2}$, beyond which contribution from
shielded low-temperature dust to the emission makes
the colour temperature lower.\footnote{{Although the dust temperature in the shielded layer
could become lower than the CMB temperature (i.e.\ the CMB heating is important),
such cold dust has a negligible impact on the colour temperature.}}
In the one-$T$ model, in contrast, the dust temperature rises monotonically
even at high $\Sigma_\mathrm{dust}$ {because of the increase in
$\Sigma_\mathrm{SFR}$ (equation \ref{eq:KS_dust})}.
%%This is because the
%%dust temperature is determined by the ratio between the total heating radiation luminosity
%%and the total dust mass; the former increases with $\Sigma_\mathrm{dust}^{1.4}$
%%(with a fixed dust-to-gas ratio) because of the KS law, while
%%the latter is just proportional to $\Sigma_\mathrm{dust}$.

We also plot the observational data of the same galaxy sample as above (Table \ref{tab:sample})
in Fig.\ \ref{fig:clr_dust}.
As already noted in Section \ref{subsec:SFR_Tdust}, lower dust-to-gas ratios or bursty star formation
activities are preferred by the data.
In particular, the upper left object in this figure (MACS0416\_Y1) is likely to be dust-poor
with $\mathcal{D}\sim 10^{-4}$ {\citep[see also][]{Sommovigo:2022aa}}.
Since this object hosts an intense star formation activity
(as shown by its high $\Sigma_\mathrm{SFR}$; Fig.\ \ref{fig:clr_SFR}) and a low dust abundance,
the dust is efficiently heated with little shielding.

%%Although the dust temperatures are similar between the RT model and the one-$T$
%%model at $\Sigma_\mathrm{dust}<10^7$ M$_{\sun}$ kpc$^{-1}$, the colour temperature saturates
%%or declines at higher dust surface densities in the RT model
%%(Fig.\ \ref{fig:clr_dust}a) for the lowest $\mathcal{D}$ ($10^{-4}$) or
%%$\mathcal{D}\sim 10^{-3}$ with $\kappa_\mathrm{s}$.
Using both Figs.\ \ref{fig:clr_SFR} and \ref{fig:clr_dust}, we could roughly infer the typical
dust-to-gas ratio of the sample. We exclude MACS0416\_Y1 already discussed above.
If the KS law holds for high-redshift galaxies, $\mathcal{D}\lesssim 10^{-4}$ is not
accepted because $\Sigma_\mathrm{SFR}$ becomes to high to be consistent with the observed
SFR surface densities.
For example, if $\mathcal{D}=10^{-4}$ and $\Sigma_\mathrm{dust}\sim 10^7$ M$_{\sun}$ kpc$^{-2}$,
where $T_\mathrm{clr}$ peaks in Fig.~\ref{fig:clr_dust}a,
the gas surface density is $\Sigma_\mathrm{gas}\sim 10^{11}$ M$_{\sun}$ kpc$^{-2}$, leading
to $\Sigma_\mathrm{SFR}\sim 2.5\times 10^3$ M$_{\sun}$ kpc$^{-2}$ from the KS law
(equation \ref{eq:KS}).
This high value is beyond the range of
observed SFR surface densities (Fig.\ \ref{fig:clr_SFR}), 
except for MACS0416\_Y1. This is why the peak of $T_\mathrm{clr}$ does not appear in the
range of $\Sigma_\mathrm{SFR}$ plotted in Fig.\ \ref{fig:clr_SFR}.
Therefore, the sample (except MACS0416\_Y1) should have $\mathcal{D}$ higher
than $10^{-4}$ if the KS law holds.
On the other hand, a high value of $\mathcal{D}\gtrsim 10^{-2}$ has difficulty in explaining the
observed high dust temperatures as argued above.
These arguments imply that the dust-to-gas ratios in $z>5$ LBGs are typically significantly lower
than $10^{-2}$ but higher than $10^{-4}$; that is, of the order of
$\mathcal{D}\sim 10^{-3}$.
{This value implies a low metallicity: if we use the $\mathcal{D}$--$Z$ relation
in nearby galaxies \citep{Remy-Ruyer:2014aa}, the above value of $\mathcal{D}$ roughly corresponds
to $Z\sim 0.2$ Z$_{\sun}$.}

Note that the above results and arguments assumed the KS law, which is poorly confirmed for
galaxies at $z\gtrsim 5$.
To avoid the conclusions being strongly affected by the assumed
KS law, we also examine the relations that do not assume a star formation law in Section \ref{subsec:wo_KS}.

\subsection{Relations without assuming the KS law}\label{subsec:wo_KS}

In this subsection, we examine how the surface densities of SFR and dust mass
determine the dust temperature without assuming the KS law
(without equation \ref{eq:KS_dust}). That is,
we treat $\Sigma_\mathrm{SFR}$ and $\Sigma_\mathrm{dust}$ as independent parameters.
In this case, we do not meed to specify $\mathcal{D}$ or $\kappa_\mathrm{s}$.

\begin{figure}
\begin{center}
\includegraphics[width=0.48\textwidth]{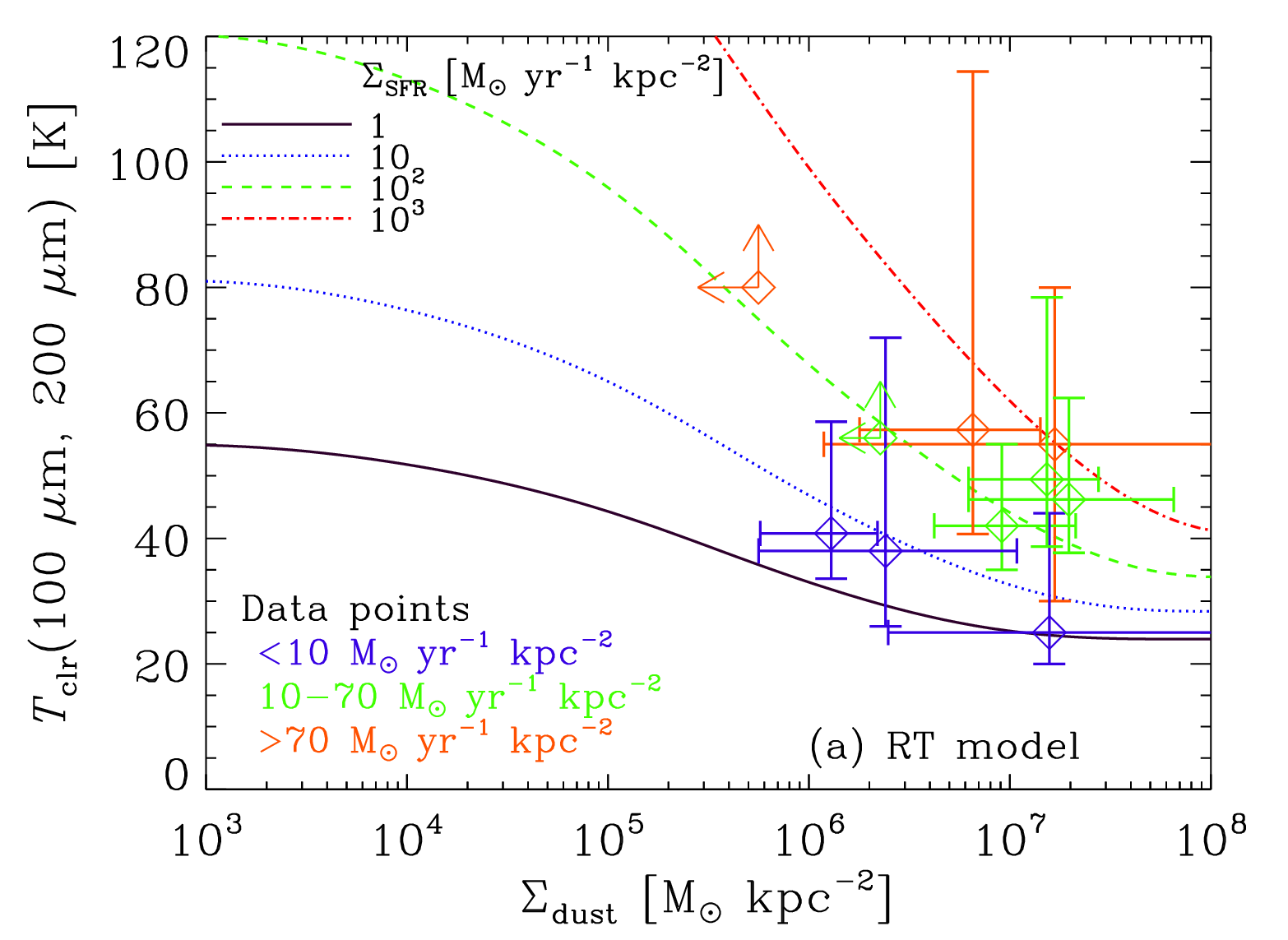}
\includegraphics[width=0.48\textwidth]{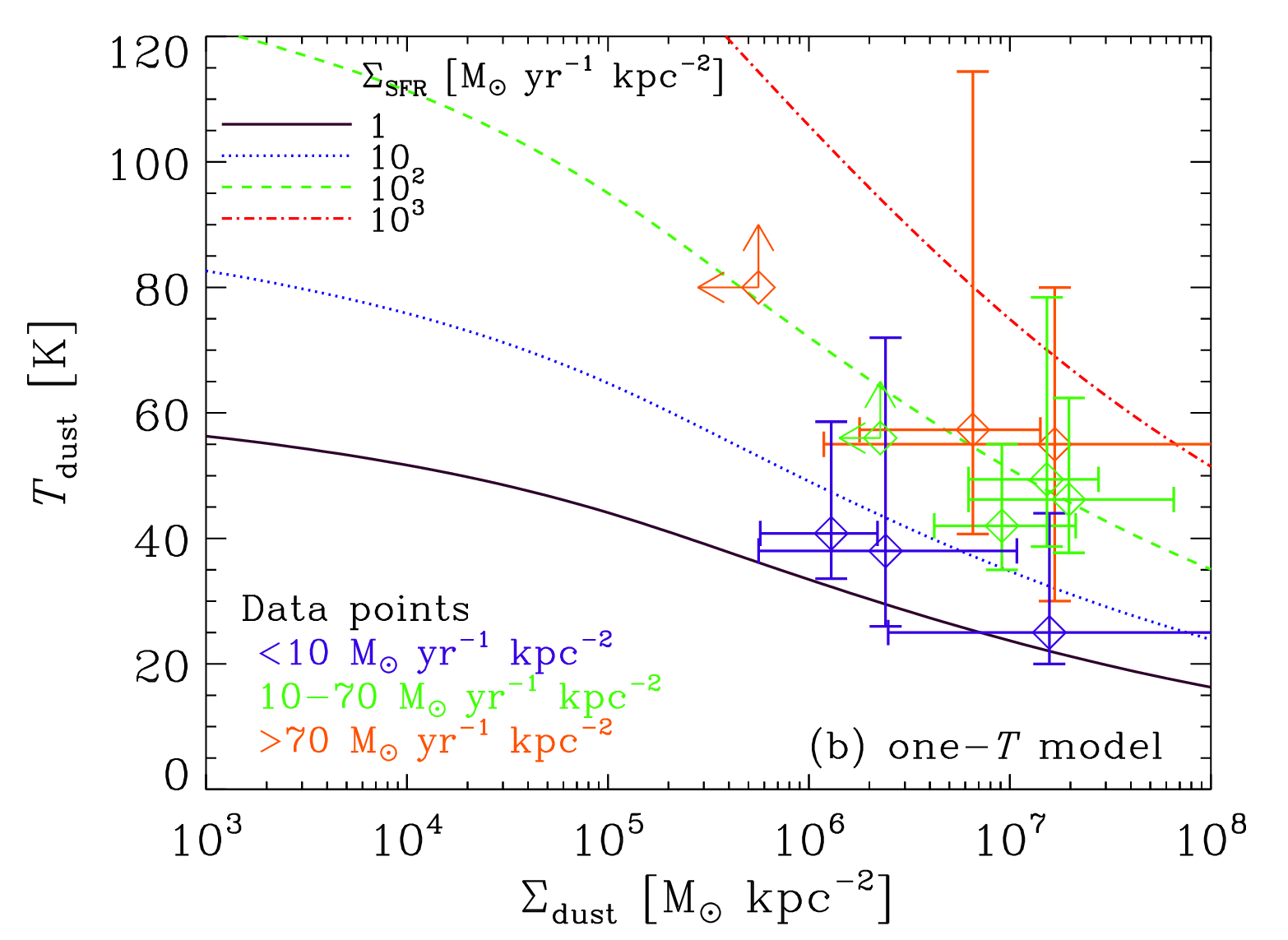}
\end{center}
\caption{Same diagrams as in Fig.\ \ref{fig:clr_dust} but with fixed
values of $\Sigma_\mathrm{SFR}$ (i.e.\ without using the KS law)
for the (a) RT and (b) one-$T$ model.
The solid, dotted, dashed, and dot--dashed lines
present the results for $\Sigma_\mathrm{SFR}=1$, 10, $10^2$ and $10^3$
M$_{\sun}$ yr$^{-1}$ kpc$^{-2}$, respectively.
The observational data points are the same as in Fig.\ \ref{fig:clr_dust} but
are colour-coded according to the SFR surface density
($<10$, 10--70, and $>70$ M$_{\sun}$ yr$^{-1}$ kpc$^{-2}$ in blue, green and red,
respectively).
\label{fig:clr_dust_freeSFR}}
\end{figure}

In Fig.\ \ref{fig:clr_dust_freeSFR}, we show the relation between dust temperature
and $\Sigma_\mathrm{dust}$ with various $\Sigma_\mathrm{SFR}$.
The dust temperature becomes lower as the dust surface density increases because
of the shielding effect.
%%The drop of dust temperature with $\Sigma_\mathrm{dust}$ is
%%steeper if the SFR surface density is higher. As a consequence, the difference in
%%the dust temperature among various $\Sigma_\mathrm{SFR}$ values shirinks as
%%$\Sigma_\mathrm{dust}$ becomes higher.
%%The shielding effect also narrows down the absolute range of dust temperature as
%%$\Sigma_\mathrm{dust}$ increases,
We observe a large difference
in the dust temperature ($\sim 30$--80 K) even at high dust surface
density ($\Sigma_\mathrm{dust}\sim 10^7$ M$_{\sun}$ kpc$^{-2}$)
for the range of $\Sigma_\mathrm{SFR}$ actually observed for high-redshift galaxies.
The range of dust temperature is consistent with the variety in $T_\mathrm{dust}$ observed
for the galaxies at $z\gtrsim 7$.

We also compare the results with the observational data in Fig.~\ref{fig:clr_dust_freeSFR}.
These data are roughly explained by
$\Sigma_\mathrm{SFR}\sim 1$--$10^3$ M$_{\sun}$ yr$^{-1}$ kpc$^{-2}$. This range is
broadly consistent with the actually observed SFR surface densities
(Fig.\ \ref{fig:clr_SFR}). For comparison,
we colour-code the observational data according to the range of
$\Sigma_\mathrm{SFR}$ ($<10$, 10--70, and $>70$ M$_{\sun}$ yr$^{-1}$ kpc$^{-2}$,
in blue, green, and orange, respectively).
The blue-green-orange trend in the observational data
is indeed consistent with the tendency of theoretical predictions with rising $\Sigma_\mathrm{SFR}$
(also shown by blue, green and orange lines). Although the large error bars hinder drawing a firm
conclusion, the overall consistency in the
$T_\mathrm{dust}$--$\Sigma_\mathrm{dust}$--$\Sigma_\mathrm{SFR}$ relation indicates the
success of our models.

Comparing the two panels in Fig.\ \ref{fig:clr_dust_freeSFR}, we find that
the two (RT and one-$T$) models are similar.
The largest difference between the two models appears at high dust surface density as also found
in Fig.\ \ref{fig:clr_dust}: In the one-$T$ model, all dust has a single temperature,
which monotonically drops as the dust mass increases. In the RT model, in contrast,
the drop of dust temperature is saturated at high $\Sigma_\mathrm{dust}$ and
low $\Sigma_\mathrm{SFR}$ because the shielded portion of dust has too low a temperature
to contribute significantly to the luminosity at $\lambda\leq 200~\micron$.
{It is reminded here that the CMB heating, which we neglected (Section \ref{sec:model}),
should be included if we are
interested in the temperature drop at high $\Sigma_\mathrm{dust}$. In particular, for
the one-$T$ model, the dust temperature would not continue to drop below the CMB
temperature towards high
$\Sigma_\mathrm{dust}$.}

The result shown in Fig.\ \ref{fig:clr_dust_freeSFR} also indicates that,
if we obtain two of the three quantities
($T_\mathrm{dust}$, $\Sigma_\mathrm{dust}$, and $\Sigma_\mathrm{SFR}$),
we can estimate the other using our model. An interesting application would be to
obtain $T_\mathrm{dust}$ from $\Sigma_\mathrm{dust}$ and $\Sigma_\mathrm{SFR}$,
both of which
could be estimated from rest-frame UV data as well (through the UV spectral index
and the UV luminosity; see section 2 of \citealt{Ferrara:2017aa}).
\citet{Ferrara:2022aa} estimated the dust temperature basically in this way.
Another application would be to derive $\Sigma_\mathrm{SFR}$ from
$T_\mathrm{dust}$ and $\Sigma_\mathrm{dust}$. If the obtained $\Sigma_\mathrm{SFR}$ is
converted to $\Sigma_\mathrm{gas}$ using the KS law, we could estimate the dust-to-gas ratio
($\Sigma_\mathrm{dust}/\Sigma_\mathrm{gas}$). We already constrained the dust-to-gas ratio
in this way in
Section \ref{subsec:Sigma_dust}, and argued that the typical dust-to-gas ratio is of the order of
$\sim 10^{-3}$.
%%If a galaxy is not resolved and only $M_\mathrm{dust}$, SFR, and $T_\mathrm{dust}$
%%are obtained, it is also possible to estimate
%%the extent of dust distribution ($R=\sqrt{ab}/2$) by searching for a set of
%%$(\Sigma_\mathrm{dust},\,\Sigma_\mathrm{SFR}=(M_\mathrm{dust}/(\upi R^2),\, \mathrm{SFR}/(\upi R^2))$
%%with the known $T_\mathrm{dust}$ that satisfies our model.
%%It is also interesting to compare these indirect methods of estimating dust temperature
%%and star formation rate (surface density)
%%with those utilizing [C \textsc{ii]} emission luminosity \citep{Sommovigo:2021aa}.

\section{Discussion}\label{sec:discussion}

\subsection{Radiative transfer and one-$T$ models}\label{subsec:radtr_vs_oneT}

In the above, we have shown that the RT and one-$T$
models overall predict similar dust temperatures. However, the difference between the two models
appears at $\Sigma_\mathrm{dust}\gtrsim 10^7$ M$_{\sun}$ kpc$^{-2}$, where the RT
model shows a saturation or decrease of the dust temperature (colour temperature)
because of shielding (Fig.\ \ref{fig:clr_dust}; Section \ref{subsec:Sigma_dust}).
%%This is because of the contribution
%%from shielded dust component, which has low $T_\mathrm{dust}\lesssim 10$ K in the
%%RT model (with a screen geometry).
In the one-$T$ model, in contrast,
the dust temperature always continues to increase even if the dust surface density increases
beyond $\Sigma_\mathrm{dust}\sim 10^7$ M$_{\sun}$ kpc$^{-2}$. 
Thus, the dust--stars distribution geometry, which affects shielding of dust-heating
radiation,
has a significant impact on the dust temperature
at $\Sigma_\mathrm{dust}\gtrsim 10^7$ M$_{\sun}$ kpc$^{-2}$. This in turn means that the
geometry of dust and star distributions only has a minor influence on the dust temperature
at lower dust surface densities.

\begin{figure}
\begin{center}
\includegraphics[width=0.48\textwidth]{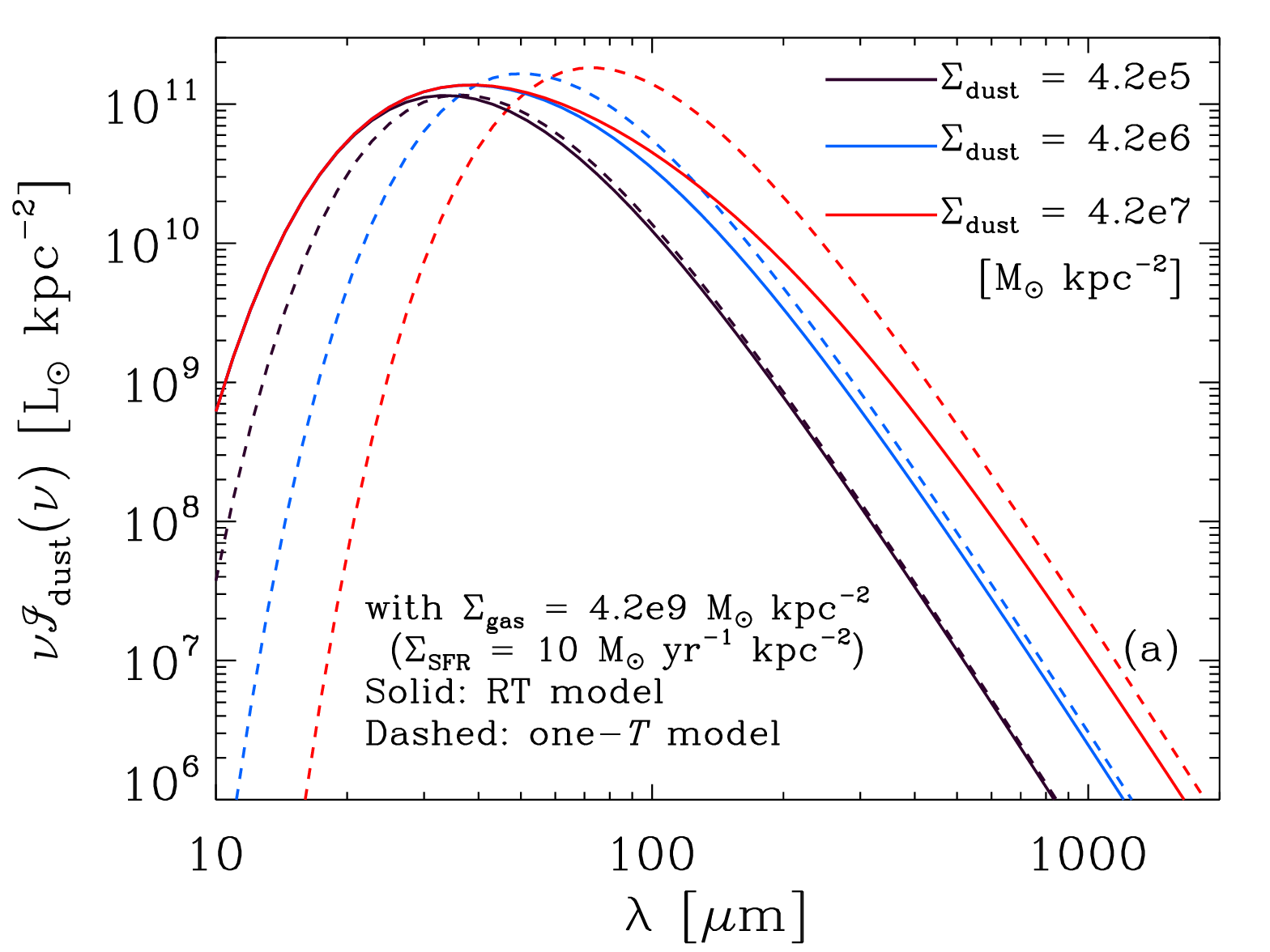}
\includegraphics[width=0.48\textwidth]{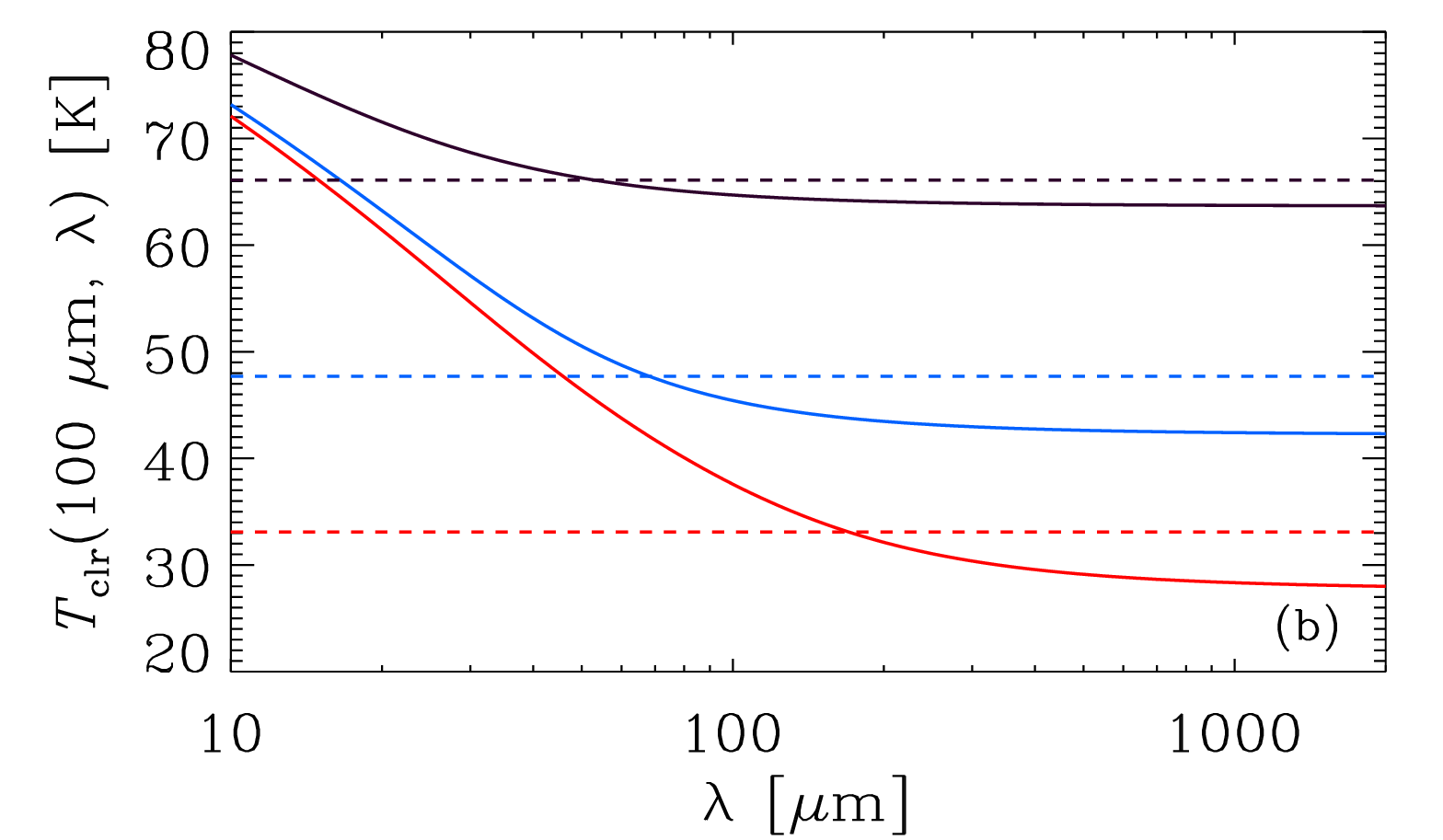}
\end{center}
\caption{(a) Dust emission SEDs for
{$\Sigma_\mathrm{gas}=4.2\times 10^9$ M$_{\sun}$ kpc$^{-2}$
(corresponding to $\Sigma_\mathrm{SFR}=10$ M$_{\sun}$ yr$^{-1}$ kpc$^{-2}$
in the KS law).
In order to examine the effect of dust optical depth (dust surface density) on the
SED, we show the results with various dust-to-gas ratios,
$\Sigma_\mathrm{dust}=4.2\times10^5$, $4.2\times 10^6$, and
$4.2\times 10^7$ M$_{\sun}$ kpc$^{-2}$ (black, blue and red lines, respectively),
corresponding to $\mathcal{D}=10^{-4}$, $10^{-3}$, and $10^{-2}$, respectively},
in the RT model (solid lines) and the one-$T$
model (dashed lines). The vertical axis shows $\Sigma_\mathrm{lum,dust}^i(\nu )$
($i=\mathrm{RT}$ or one-$T$ depending on the model) multiplied by the frequency.
(b) Colour temperature as a function of wavelength for the RT model (solid lines).
One of the wavelengths is fixed to 100 $\micron$, and the other is moved freely.
For comparison, we also show the dust temperature in the one-$T$ model
(dashed lines). Each colour corresponds to the same value of $\Sigma_\mathrm{dust}$ as
in Panel (a).
\label{fig:SED_comp}}
\end{figure}

From the above results, we expect that the SED shapes of dust emission are different between the two
(RT and one-$T$) models at high dust {surface} densities.
To visualize this expectation, we present in Fig.~\ref{fig:SED_comp}
the SEDs for various $\Sigma_\mathrm{dust}$ with a fixed $\Sigma_\mathrm{gas}$
(so a fixed $\Sigma_\mathrm{SFR}$, whose value is determined by the KS law).
We choose the case of $\Sigma_\mathrm{SFR}=30$ M$_{\sun}$ yr$^{-1}$ kpc$^{-2}$
($\Sigma_\mathrm{gas}\simeq 4.2\times 10^9$ M$_{\sun}$ kpc$^{-2}$), which
is roughly in the middle of the observational sample we adopted (Fig.~\ref{fig:clr_SFR}).
We examine
$\Sigma_\mathrm{dust}=4.2\times 10^5$, $4.2\times 10^6$, and $4.2\times 10^7$ M$_{\sun}$ kpc$^{-2}$,
which correspond to $\mathcal{D}=10^{-4}$, $10^{-3}$, and $10^{-2}$, respectively.

Note that, in reality, the emission at wavelengths well below the SED peak position is not precisely
predicted in our model, since stochastically heated very small grains
\citep[e.g.][]{Draine:1985aa}, which are not included in our model, contribute to the emission
significantly.
{Moreover, complex dust--stars geometries on small spatial scales, which are
not included in our models (see Section \ref{subsec:complex}) would
also lead to hot dust components located in the vicinity of compact, actively star-forming regions.
Such hot dust components contribute to emission at short wavelengths
\citep{Sommovigo:2020aa}, which is missing in
our prediction. Such compact region could enhance the optical depth, and possibly make the
region optically thick for short-wavelength dust emission.}
Thus, we do not discuss the difference in SED shape on the Wien side, but focus on
the wavelengths around the SED peak and on the Rayleigh--Jeans side.

We observe in Fig.\ \ref{fig:SED_comp} that the two (RT and one-$T$) models
show different trends with increasing $\Sigma_\mathrm{dust}$. In the RT model,
the SED extends to longer wavelengths as the dust abundance becomes larger with the luminosity
at the shortest wavelengths fixed. This is because cold layers of dust are added if we increase
$\Sigma_\mathrm{dust}$ with a fixed value of $\Sigma_\mathrm{SFR}$ (i.e.\ a fixed stellar luminosity).
In contrast, in the one-$T$ model, the SED shifts towards longer wavelengths as
$\Sigma_\mathrm{dust}$ increases, reflecting the drop of dust temperature. This is because
the stellar radiation received per dust mass decreases as the dust increases.
In both models, we see a slight rise of the SED peak with $\Sigma_\mathrm{dust}$
simply because of the increase in
the energy absorbed by dust.
At low $\Sigma_\mathrm{dust}$, the
difference between the two models is small. This means that the dust temperature is well approximated with
a single value in the RT model because {shielding} is weak.
In contrast, at high $\Sigma_\mathrm{dust}$,
the SEDs are different between the two models.
Therefore, if the dust surface density is as high as $\Sigma_\mathrm{dust}\gtrsim 10^7$
M$_{\sun}$ kpc$^{-2}$, detailed dust temperature structures created by RT
effects are important in the detailed SED shape.

From the difference in SED shape between the two models,
we expect that the colour temperature in the RT model depends on the selected
wavelengths at high $\Sigma_\mathrm{dust}$. In Fig.\ \ref{fig:SED_comp}b, we show the
colour temperature as a function of wavelength. We fix one of the wavelengths to 100 $\micron$
and move the other freely, and present $T_\mathrm{clr}(100~\micron ,\,\lambda )$.
We observe that the colour temperature monotonically decreases as $\lambda$ increases because
we selectively observe lower-temperature dust at longer wavelengths.
If we focus on long (rest-frame) wavelengths
($\lambda\gtrsim 100~\micron$), which are often used by ALMA observations
(Bands 6--8) of high-redshift galaxies,
the colour temperature is not sensitive to the selected wavelengths.
When we use Band 9 (450 $\micron$) for galaxies at $z\gtrsim 7$
(i.e.\ $\lambda\lesssim 60~\micron$),
the colour temperature is systematically high {in the RT model}
since high-temperature layers dominate the emission
at such a short wavelength. Thus, dust temperature estimates including Band 9 need to be
carefully interpreted by noting {a possibility of} multi-$T_\mathrm{dust}$ structures.
At long wavelengths, such a
multi-temperature effect is not important; indeed,
the colour temperature is almost constant at $\lambda\gtrsim 200~\micron$, and
is similar to the dust temperature in the one-$T$ model.
Thus, the above predictions on the colour temperature are not altered significantly
as long as we focus on ALMA Bands 6--8.
{If we include a Band 9 observation in the SED analysis, it is safer to
include multiple bands from Bands 6--8 as well in order to examine the multi-$T_\mathrm{dust}$
effect on the SED.}

\subsection{Effects of dust properties}\label{subsec:dust_properties}

\begin{figure}
\begin{center}
\includegraphics[width=0.48\textwidth]{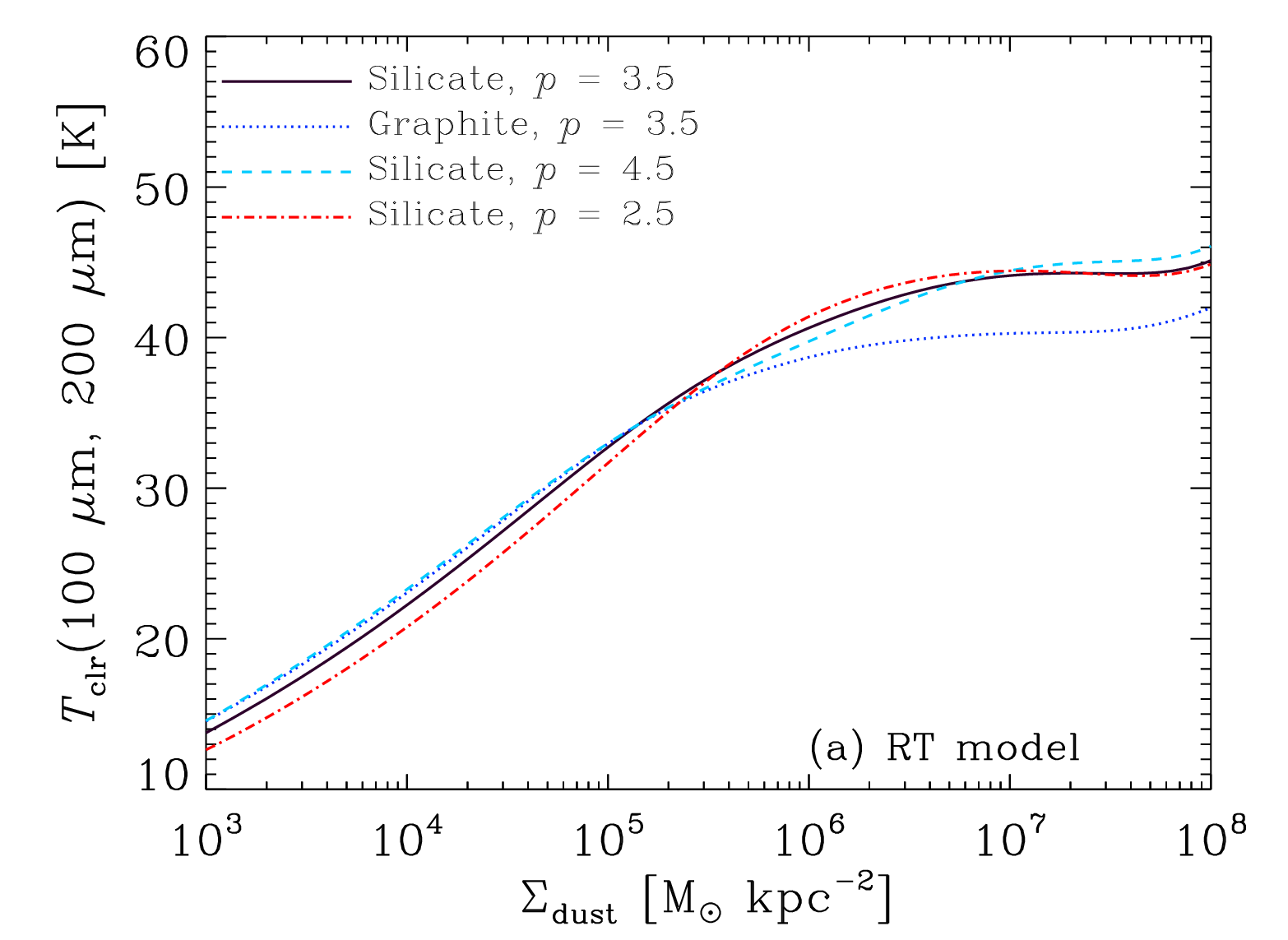}
\includegraphics[width=0.48\textwidth]{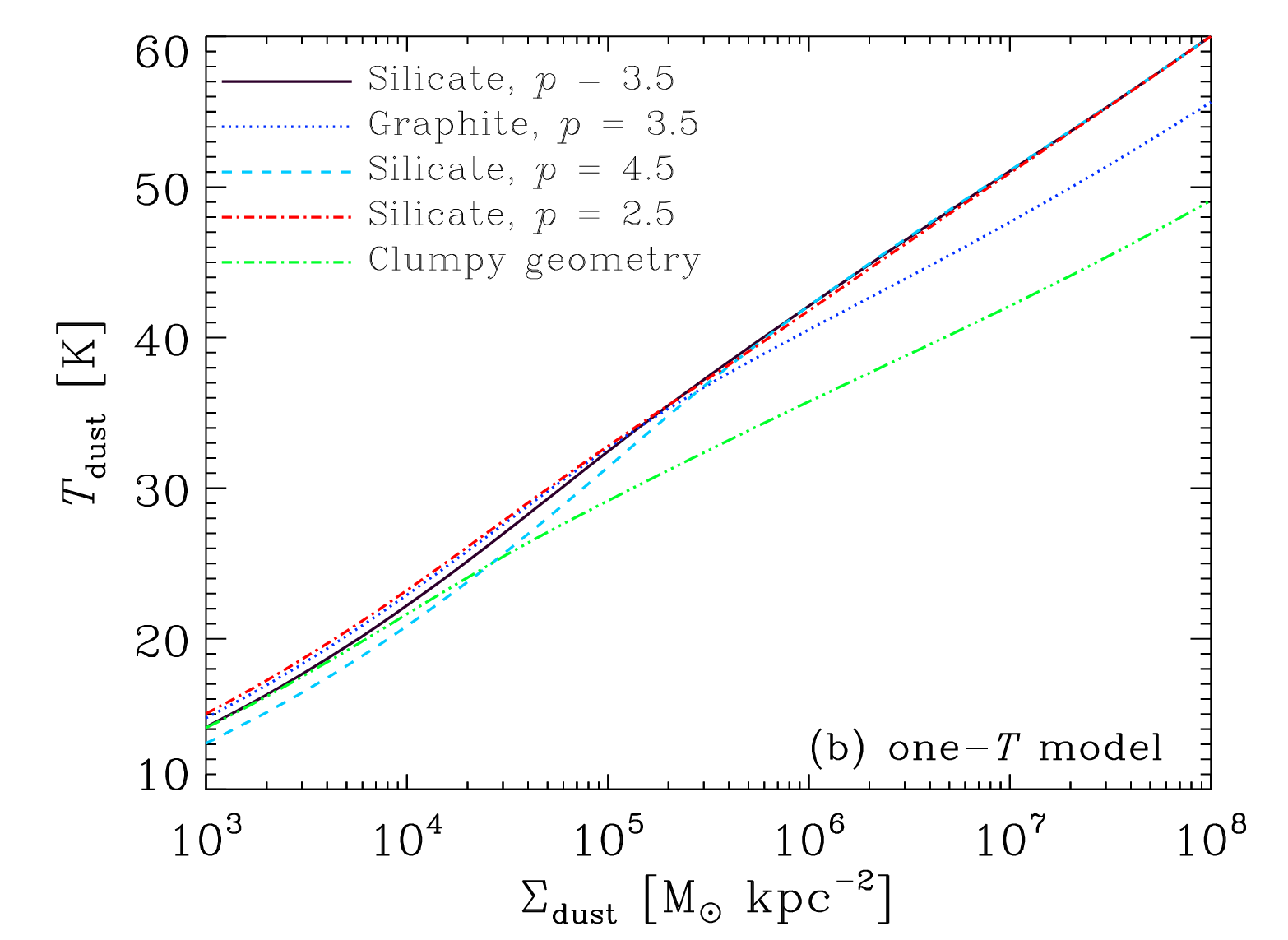}
\end{center}
\caption{Dust temperature as a function of dust surface density in the (a) RT
and (b) one-$T$ models
for a variety of dust properties. We adopt $\mathcal{D}=10^{-3}$ and $\kappa_\mathrm{s}=1$.
The solid, dashed, and dot--dashed lines
correspond to silicate with $p=3.5$ (MRN), 4.5, and 2.5, respectively. The dotted line shows graphite with
$p=3.5$. The homogeneous geometry is adopted for the one-$T$ model.
In Panel (b), we also show the result of the clumpy geometry with $\xi_\mathrm{cl}=3$ for
silicate with $p=3.5$ (triple-dot--dashed line) in addition
to the honogeneous geometry.
\label{fig:dust_properties}}
\end{figure}

We examine the variation of dust properties. We vary the dust composition and
the grain size distribution (Section \ref{subsubsec:dust_properties}).
We fix $\mathcal{D}=10^{-3}$. We show the relation between dust temperature and
$\Sigma_\mathrm{dust}$ for silicate and graphite with $p=3.5$, and for $p=2.5$ and 4.5 with
silicate in Fig.\ \ref{fig:dust_properties}. The burstiness parameter is fixed to $\kappa_\mathrm{s}=1$.

We observe in Fig.\ \ref{fig:dust_properties} that the dust temperature is insensitive to
$p$ in both RT and one-$T$ models.
While large $p$ indicates more efficient absorption of UV radiation (because of more small grains),
it also means
more efficient shielding. These two effects counteract each other.
The difference between silicate and graphite is more
apparent, especially at high $\Sigma_\mathrm{dust}$. This is not only due to more efficient shielding,
but also because of more efficient emission (higher mass absorption coefficient) for graphite.
More efficient emission leads to a lower equilibrium dust temperature.
Comparing the RT and one-$T$ models, we confirm significant difference
at high $\Sigma_\mathrm{dust}$ as pointed out above.
%%This is due to more efficient shielding in the
%%RT model.
In both models, graphite predicts lower dust temperatures than silicate
at dust {surface} densities ($\sim 10^7$ M$_{\sun}$ kpc$^{-2}$) appropriate for
the $z>7$ sample above
but the difference is only $\sim 10$ per cent.

In the one-$T$ model (Fig.\ \ref{fig:dust_properties}b),
we also show the clumpy geometry with
$\xi_\mathrm{cl}=3$
(for silicate with $p=3.5$), which shows significantly lower dust temperature than the homogensous geometry
at high $\Sigma_\mathrm{dust}$. This is because, as explained in Section \ref{subsubsec:oneT},
dust only covers the galaxy surface with a fraction of $1/\xi_\mathrm{cl}$. This means that
dust can only absorb a part of stellar light even if the dust abundance is high.
Since the total emission energy of dust is scaled with $T_\mathrm{dust}^{4+\beta}$, we obtain
a scaling of $T_\mathrm{dust}\propto\xi_\mathrm{cl}^{-1/(4+\beta)}$
($\xi_\mathrm{cl}>1$) at high dust surface density. This scaling is useful to infer the dust temperature
with different values of $\xi_\mathrm{cl}>1$ (recall again that the result is similar to the homogensous case
with $\xi_\mathrm{cl}\lesssim 1$) at high dust surface densities.
At low dust surface densities, the dust is optically thin for UV radiation; in this case, the total radiation energy
absorbed by dust is determined by the total dust mass, and is not sensitive to dust distribution geometry.
Thus, clumpy geometry gives a conservative (low) dust temperature, which means that
the requirement for low $\mathcal{D}$
or higher $\kappa_\mathrm{s}$ is pronounced if we aim at explaining the high
dust temperatures with clumpy geometry compared with the uniform geometry.

\subsection{Further complexities}\label{subsec:complex}

Although the geometries of dust--stars distribution in real galaxies are complex,
we still expect that our theory based on the surface densities are applicable to
a variety of galaxies. This is because the radiation field is also a
`surface' quantity in the sense that it has the same physical dimension as the
surface luminosity (Section \ref{subsubsec:galaxy}).
Our predictions are further checked with observations of
nearby galaxies (Chiang et al., in preparation). Numerical simulations with
more complex dust--stars geometries could also be used to examine the robustness
of our predictions.

However, the effect of local intense radiation sources may not be
included in our simple treatment. Some analytic and numerical studies showed
that a dust component concentrated nearby an intensely star-forming region could
have a large contribution to the FIR luminosity of the galaxy
\citep{Behrens:2018aa,Sommovigo:2020aa,Pallottini:2022aa}.
Complex dust--stars geometries are also shown observationally for galaxies
at $z\sim 7$ \citep{Willott:2015aa,Bowler:2022aa}.
Therefore, our simple analytic treatments should be carefully applied to
real high-redshift galaxies, and studies focusing on small-scale structures should
supplement our understanding of what regulates the dust temperature in high-redshift
galaxies. Interestingly, the necessity of low dust abundance in explaining the high
dust temperature is common between our work and some studies that included small-scale or
complicated geometries \citep{Liang:2019aa,Ma:2019aa,Sommovigo:2022aa}.

\section{Conclusions}\label{sec:conclusion}

For the purpose of interpreting observed dust temperatures at high redshift
($z>5$),
we construct analytic models to calculate the dust temperature under given
star formation activity and dust properties (especially dust abundance).
The models are described by
the surface densities of gas mass and SFR since the surface quantities are
important to describe the radiation field intensity and the dust optical depth.
We develop the following two models that
can be treated analytically: (i) RT and (ii) one-$T$ models.
In the first model, we {consider} the multi-temperature (or dust shielding) effect
{within the framework of plane-parallel treatment} by
putting the stars in the midplane of the disc and the dust in the screen geometry.
The dust temperature in this model is defined by the
colour temperature at two selected wavelengths (100 and 200 $\micron$ by default).
In the second model, we consider an opposite extreme by considering that dust
and stars are well mixed, so that the dust is assumed to have a single temperature.
In this model, the dust temperature is determined by the global balance between
the absorbed and radiated energy by the dust.
{These two extremes serve to bracket the most realistic scenario}.
%%We expect that these two models `sandwich' realistic situations in terms of the dust temperature.

We particularly focus on the dust temperature as a function of
SFR surface density ($\Sigma_\mathrm{SFR}$) and dust surface density
($\Sigma_\mathrm{dust}$).
%%Observational data are also available for these quantities at $z>5$.
As expected, the dust temperature rises with
increasing $\Sigma_\mathrm{SFR}$ and $\Sigma_\mathrm{dust}$ (which has
a positive relation with $\Sigma_\mathrm{SFR}$ because of the KS law).
However, these relations depend on the dust-to-gas ratio ($\mathcal{D}$), since it affects the
relation between $\Sigma_\mathrm{SFR}$ and $\Sigma_\mathrm{dust}$
(equation \ref{eq:KS_dust}).
Lower values of $\mathcal{D}$ predicts higher dust temperatures.
Thus, low dust abundance ($\lesssim 10^{-3}$) can be a reason for observed high dust temperatures
{($T_\mathrm{dust}\gtrsim 40$ K)}
in high-redshift galaxies. Another reason could be a burst of star formation (i.e.\ high $\kappa_\mathrm{s}$).
The grain size distribution and the dust composition have less
impacts on the dust temperature than
$\mathcal{D}$ and $\kappa_\mathrm{s}$.

The RT and one-$T$ models predict similar dust temperatures
except at high dust surface density ($\Sigma_\mathrm{dust}>10^7$ M$_{\sun}$ kpc$^{-2}$).
Some ALMA-detected galaxies at $z>5$ may be
located in this high-$\Sigma_\mathrm{dust}$ regime, which means that a careful
radiative transfer treatment is necessary to predict precise dust temperature.
However, the difference among different values of $\mathcal{D}$ and $\kappa_\mathrm{s}$
is significant, and the conclusion that high-redshift LBGs favour low
$\mathcal{D}\lesssim 10^{-3}$ (if $\kappa_\mathrm{s}\lesssim 10$) is not altered.

We also examine the relation between dust temperature and $\Sigma_\mathrm{dust}$
without assuming the KS law; that is, we treat $\Sigma_\mathrm{SFR}$ as an independent parameter.
Overall, higher $\Sigma_\mathrm{SFR}$ indicates higher dust temperature
and we predict $T_\mathrm{dust}\sim 30$--80 K
for the range of $\Sigma_\mathrm{SFR}$ and $\Sigma_\mathrm{dust}$
appropriate for high-redshift ($z>5$)
LBGs. The observational data ($\Sigma_\mathrm{SFR}$, $\Sigma_\mathrm{dust}$,
and $T_\mathrm{dust}$) of $z>5$ LBGs are consistent with
the calculation results. Interestingly, we also find a trend that LBGs with higher
$\Sigma_\mathrm{SFR}$ and lower $\Sigma_\mathrm{dust}$ tend to have higher $T_\mathrm{dust}$,
which is consistent with our prediction (Fig.\ \ref{fig:clr_dust_freeSFR}).

The difference between the two (RT and one-$T$) models is further
examined. We observe a significant difference in SED shape between the two models at
$\Sigma_\mathrm{dust}\gtrsim 10^7$ M$_{\sun}$ kpc$^{-2}$ since the superposition of
layers with various dust temperatures is important at such a high dust surface density
in the RT model.
Thus, if the dust surface density is higher than $\sim 10^7$ M$_{\sun}$ kpc$^{-2}$,
a detailed radiative transfer calculation is necessary to discuss the detailed shape of
dust emission SED. We, however, find that, in the range of dust surface density appropriate for
high-redshift LBGs, the colour temperature in the RT model is
similar to the dust temperature in the one-$T$ model as long as we use
$\lambda\sim 100$--200 $\micron$.
Thus, the dust temperature measured in ALMA Bands 6--8 ($\sim 650$--1,200 $\micron$)
for $z\gtrsim 5$ galaxies
is not sensitive to the detailed
radiative transfer effects. Note that Band 9 ($\sim 450~\micron$) may selectively
observe high-dust-temperature layers {at
$\Sigma_\mathrm{dust}\gtrsim 10^7$ M$_{\sun}$ kpc$^{-2}$, if the multi-$T_\mathrm{dust}$
structure is as significant as realized in the RT model.}

In the one-$T$ model, we also investigate the effect of clumpiness in dust distribution geometry.
We only consider the case where the density contrast between the clumps and the diffuse medium is
large since otherwise the resulting dust temperature is similar to the homogeneous geometry.
At low $\Sigma_\mathrm{dust}$, the clumpy geometry predicts almost the same dust temperature
as the homogeneous geometry. However, at high $\Sigma_\mathrm{dust}\gtrsim 10^7$ M$_{\sun}$ kpc$^{-2}$,
the dust temperature is lower in the clumpy case because
the dust effectively covers only a certain fraction of the galaxy surface.
This strengthens the requirement of low dust-to-gas ratio and/or high $\kappa_\mathrm{s}$
to achieve a high dust temperature.

From the above, we conclude that the high dust temperatures
{($T_\mathrm{dust}\gtrsim 40$ K)}
in {some} ALMA-detected
$z\gtrsim 7$ galaxies is caused by a low dust-to-gas ratio
($\mathcal{D}\lesssim 10^{-3}$) if the KS law holds in high-redshift galaxies.
A burst-like star formation with $\kappa_\mathrm{s}\gtrsim 10$ could give another
explanation for the high dust temperatures.
These conclusions are not sensitive to the dust properties (dust composition and
grain size distribution) and detailed radiative transfer effect.
In the companion paper (Chiang et al., in preparation), we test our model using
spatially resolved observations of nearby star-forming galaxies.

\section*{Acknowledgements}
 
{We are grateful to the anonymous referee for useful comments.}
HH thanks the Ministry of Science and Technology (MOST) for support through grant
MOST 108-2112-M-001-007-MY3, and the Academia Sinica
for Investigator Award AS-IA-109-M02.

\section*{Data Availability}

Data related to this publication and its figures are available on request from the corresponding author.



%\bibliographystyle{mnras}
\bibliographystyle{mnras}
\bibliography{/Users/hirashita/bibdata/hirashita}
%%\bibliography{hirashita}

%%%%%%%%%%%%%%%%% APPENDICES %%%%%%%%%%%%%%%%%%%%%
%%\appendix

%%\section{To be deleted from the submitted version}\label{app:derivation}

% Don't change these lines
\bsp	% typesetting comment
\label{lastpage}
\end{document}